# Additional contributions to elastic energy of lipid membranes: Tilt-curvature coupling and curvature gradient

Konstantin V. Pinigin*, Peter I. Kuzmin, Sergey A. Akimov, Timur R. Galimzyanov*

A.N. Frumkin Institute of Physical Chemistry and Electrochemistry, Russian Academy of Sciences, 31/4 Leninskiy prospekt, Moscow 119071, Russia;

*corresponding authors: pinigin@phystech.edu; gal_timur@yahoo.com

ABSTRACT

Lipid bilayer membranes under biologically relevant conditions are flexible thin laterally fluid films consisting of two unimolecular layers (monolayers) each about 2 nm thick. On spatial scales much larger than the bilayer thickness, the membrane elasticity is well determined by its shape. The classical Helfrich theory considers the membrane as an elastic two-dimensional (2D) film, which has no particular internal structure. However, various local membrane heterogeneities can result in a lipids tilt relative to the membrane surface normal. On the basis of the classical elasticity theory of 3D bodies, Hamm and Kozlov [Eur. Phys. J. E 3, 323 (2000)] derived the most general energy functional, taking into account the tilt and lipid monolayer curvature. Recently, Terzi and Deserno [J. Chem. Phys. 147, 084702 (2017)] showed that Hamm and Kozlov's derivation was incomplete because the tilt-curvature coupling term had been missed. However, the energy functional derived by Terzi and Deserno appeared to be unstable, thereby being invalid for applications that require minimizations of the overall energy of deformations. Here, we derive a stable elastic energy functional, showing that the squared gradient of the curvature was missed in both of these works. This change in the energy functional arises from a more accurate consideration of the transverse shear deformation terms and their influence on the membrane stability. We also consider the influence of the prestress terms on the stability of the energy functional, and we show that it should be considered small and the effective Gaussian curvature should be neglected because of the stability requirements. We further generalize the theory, including the stretching-compressing deformation modes, and we provide the geometrical interpretation of the terms that were previously missed by Hamm and Kozlov. The physical consequences of the new terms are analyzed in the case of a membrane-mediated interaction of two amphipathic peptides located in the same monolayer. We also provide the expression for director fluctuations, comparing it with that obtained by Terzi and Deserno.

# I. INTRODUCTION

The amphiphilic nature of lipid molecules leads to their self-assembly into bilayer lipid membranes under certain conditions in aqueous solutions. The membrane outer surfaces are hydrophilic, while the membrane interior is hydrophobic [1,2]. Bilayer lipid membranes are common for many living organisms, where they constitute the structural basis of plasma membranes, secretory vesicles, Golgi apparatus, lysosomes, etc. [3]. In biological membranes the lipid bilayer serves two main purposes: it acts as the weakly permeable barrier between the cell interior and the environment and plays the role of the matrix or platform for membrane proteins, which mediate various cell functions. The functioning of living organisms demands reshaping and topology changes of membrane structures. Intermediate states of such processes as endo- and exocytosis or various types of fusion and fission include strongly bent membranes. In some processes, the energy of lipid membrane deformations is believed to be the rate-limiting factor [4–6]. This motivates the development of methods for the analysis of membrane reshaping energetics.

Most lipid membranes under biologically relevant conditions are flexible laterally fluid films consisting of two unimolecular layers (monolayers) each about 2 nm thick. On spatial scales much larger than the bilayer thickness, the membrane elastic energy is well determined by its shape. The classical Helfrich theory considers membrane as an elastic 2D film, which has no particular internal structure [7]. However, such a large scale approach is insufficient for an adequate description of various membrane processes, especially accompanied by alteration of membrane topology. Besides, the Helfrich's approach is poorly applicable for the description of deformations induced by shallow membrane inclusions, phase boundaries, etc. To analyze these phenomena, the internal structure of membranes should be taken into account by the introduction of, for example, tilt deformation arising when the average direction of lipid tails deviates from the normal vector to the lipid monolayer surface. The tilt can appear at lipid domain boundaries [8,9], at the edge of membrane pores [10] or the boundary of membrane inclusions [11]. This deformation is exceptionally critical in processes involving alteration of membrane topology, such as fission and fusion [4,6,12,13]. In addition, lipid tilting can influence membrane fluctuations [14–17].

Various geometrical phenomenological theories adopting a two-dimensional approach and addressing the tilt degree of freedom have been proposed [7,18–26]. Many of these theories ([18,21,22,24–26]) focus on particular lamellar phases with 2D ordering of spontaneously tilted lipids, while others address the lamellar liquid crystalline phase ($L_\alpha$), which is more relevant for biological membranes and considered in this paper. The drawback of such phenomenological theories is that they are unaware of (1) how various energy contributions originate from underlining microscopic physics of lipids and (2) whether these contributions in fact take place. Actually, lipid membranes are 3D objects rather than 2D ones as they are usually considered. One of the earliest works adapting a three-dimensional approach to lipid membranes was developed by M. Hamm and M. M. Kozlov [27]. Based on the three-dimensional elasticity theory, Hamm and Kozlov (hereafter HK) wrote the most general classical quadratic expression for the elastic energy of a monolayer and took into account the following primary features of lipid monolayers: (1) the volume incompressibility, (2) the in-plane fluidity, (3) the transverse isotropy in a flat configuration and (4) the presence of a transmonolayer stress profile (also called pressure profile or pre-stress). In HK's theory, the energy of a monolayer includes an effective mean curvature, Gaussian curvature and tilt terms, the elastic moduli of which are expressed through microscopic membrane characteristics. Thus, HK theory yields a quadratic energy functional of the membrane, that upgrades the curvature-based Helfrich functional in two major aspects: i) the tilt field emerges; ii) the divergence of the tilt field contributes to the mean membrane curvature, leading to the new effective curvature field decoupled from the tilt.

There are also other three-dimensional theories of lipid membranes, which address the tilt deformation mode: the reduction from three dimensions to two of the classical Frank theory of liquid crystals [28] and the opposing forces model [16,29,30]. However, both of them do not address the transverse isotropy of lipid monolayers, which may potentially reflect some of their elastic properties. In fact, as we show in this paper, the new substantial terms arise from transverse shear deformations, which directly reflect the transverse isotropy and therefore are not present in the aforementioned theories.

HK model is widely used for the description of different membrane phenomena: a fusion [12,31–34], fission [5,6], poration [35–37], phase coexistence and phase boundary energy [38–42], and interaction of membrane inclusions with raft boundaries [43–45]. The model served as a rational basis for the development of the theory of elasticity of bolalipid membranes [46]. However, recently, M. M. Terzi and M. Deserno (hereafter TD) revisited the HK theory [47] and raised the concern about the validity of its derivation and the final expression for the elastic energy. Initial TD's derivation [47] was further improved in Ref. [48]. TD showed that the existing form of the elastic theory is incomplete, because the coupling term between the tilt and effective curvature had been missed in the original HK energy functional. However, TD's functional is unstable, which follows from a divergence of fluctuation spectra [47]. Thus, it is invalid for applications, which require minimizations of the overall energy of deformations and it is unclear to what physical consequences new energy contributions might lead. Therefore, the motivation of this paper is to find out the reasons for the instability, derive the most general stable energy functional suitable for various applications, and investigate the physical consequences of the new energy contributions.

Here, we show that the elastic energy functional should be further appended by one more term — the squared gradient of the effective curvature. We also consider the influence of the pre-stress terms on the stability of the energy functional and demonstrate that the effective Gaussian curvature should be neglected both because of stability requirements and the exceedance of the quadratic order of smallness. In addition, we figure out the physical consequences of the new terms, using the example of a membrane-mediated interaction of two amphipathic peptides located in the same monolayer of a lipid membrane. The comparison of the theoretical expression for director fluctuations with that of TD is also provided.

In Sec. II we recall the basic concepts from HK theory, including definitions of the director and tilt and some fundamental equations. We write down the expression for the elastic energy density and discuss the influence of the pre-stress on membrane stability. In Sec. III we express the terms of the energy functional via the curvature and tilt. The revisiting of the transverse shear deformation terms is also provided. In Sec. IV we discuss the order of smallness of the new terms as well as the instability issues caused by the effective Gaussian curvature. We also discuss the physical implications of new terms and their influence on membrane fluctuation spectra.

## II. FUNDAMENTAL EQUATIONS

### A. Basic notations

In this work, we follow HK's approach [27] and utilize the classical theory of elasticity [49] in order to derive the free energy functional for the lipid monolayer. We introduce a field of unit vectors $\mathbf{n}$ called directors. The field is defined at the dividing surface that separates hydrophobic and hydrophilic parts of lipids. Directors are assumed to be directed into the hydrophobic part of the monolayer; they characterize the average orientation of lipid tails. The measure of the director deviation from the unit normal $\mathbf{N}$ to the dividing surface is described by the tilt vector $\mathbf{T} = \dfrac{\mathbf{n}}{\mathbf{n} \cdot \mathbf{N}} - \mathbf{N}$, which is parallel to the dividing surface, as $\mathbf{T} \cdot \mathbf{N} = 0$. The vector fields of $\mathbf{n}$ and $\mathbf{T}$ are used to parameterize the elastic energy of the lipid monolayer. In the reference configuration, the monolayer is assumed to be flat with the directors being perpendicular to the flat dividing surface. We introduce a Cartesian coordinate system $xyz$, the $xy$-plane of which coincides with the dividing surface and the $z$-axis is directed into the hydrophobic part of the monolayer. In the reference configuration, the dividing surface is planar; its shape is given by $\mathbf{X}_0(x,y) = (x,y,0)$, and the unit normal vector by $\mathbf{N}_0 = (0,0,1)$. In this configuration, the points inside the monolayer can be parameterized as $\mathbf{X}'_0(x,y,z) = \mathbf{X}_0(x,y) + z\mathbf{N}_0$. Now, we consider an arbitrary deformation of the monolayer. In the new configuration, the coordinates of the points inside the monolayer can be described by function $\mathbf{X}'(x,y,z)$, which is parameterized as $\mathbf{X}'(x,y,z) = \mathbf{X}(x,y) + \zeta(x,y,z)\mathbf{n}(x,y)$, where function $\mathbf{X}(x,y)$ describes the shape of the deformed dividing surface and $\zeta(x,y,z)$ denotes the distance between the monolayer points and the dividing surface, measured along the director $\mathbf{n}(x,y)$. It is worth mentioning that in HK's theory $\zeta(x,y,z)$ denotes the projection of this distance to the normal to the dividing surface. However, we will use our notation, as it is convenient and does not make much of a difference. Actually, we assume that any line segment $\mathbf{r}_0(x,y,z)$ perpendicular to the dividing surface at the point $(x,y,0)$ in the reference configuration transforms upon a deformation to a curve $\mathbf{r}(x,y,\zeta)$, and the transformation can be written as a power series in its natural parameterization $\mathbf{X}'(x,y,z) = \mathbf{X}(x,y) + \zeta\mathbf{n}(x,y) + \sum_{n=2}^{\infty} \frac{\zeta^n}{n!} \frac{\partial^n \mathbf{r}}{\partial \zeta^n}(x,y,0)$, where $\zeta = \zeta(x,y,z)$ is the length of a curve part lying between the dividing surface and point $\mathbf{X}'(x,y,z)$. Hereafter, we consider only the first two terms of the expansion, i.e. $\mathbf{X}(x,y) + \zeta\mathbf{n}(x,y)$. This parameterization is also used by TD and HK [27,47].

B. Elastic energy functional

Denoting the Jacobian matrix of the function $\mathbf{X}'$ by $\nabla \mathbf{X}'$, we can write the expression for the Green-Lagrange strain tensor $\mathbf{U}$ as $\mathbf{U} = \frac{1}{2}\left( (\nabla \mathbf{X}')^T (\nabla \mathbf{X}') - \mathbf{I} \right)$, where $\mathbf{I}$ is a unit matrix. Physically, $\mathbf{U}$ represents a measure of the distance change between the points inside a body. If we consider in an arbitrary point three small intersecting line segments perpendicular to each other and parallel to the coordinate axes, then the diagonal elements of $\mathbf{U}$ will be a measure of the length change of these segments, while non-diagonal — a measure of the angle change between the segments [50]. If no deformation occurs, then $\mathbf{U}$ is a zero matrix. If we assume that all elements $u_{ij}$ of $\mathbf{U}$ are small and the monolayer is transversely isotropic in the reference configuration, then according to [27] the free energy density of the volume element is given by:

$$F = \sigma_l (u_{xx} + u_{yy}) + \sigma_{zz} u_{zz} +$$
$$+ 2\lambda_1 (u_{xx} + u_{yy})^2 + \frac{1}{2}\lambda_2 u_{zz}^2 + 2\lambda_3 (u_{xx} + u_{yy}) u_{zz} + \qquad (1)$$
$$+ \lambda_4 \left( (u_{xx} + u_{yy})^2 - 4(u_{xx} u_{yy} - u_{xy}^2) \right) + 4\lambda_5 (u_{xz}^2 + u_{yz}^2),$$

where $\lambda_i$ are the elastic moduli, which generally depend on z, and $\sigma_l, \sigma_{zz}$ — the pre-existing lateral and transverse stress inside the monolayer. This equation can be derived from the general expression for the deformation energy after applying the symmetry requirements to the Taylor series expansion up to the second order in $u_{ij}$ [51].

In general, for physically stable systems the energy functional should be positive semi-definite up to some constant, i.e. it should be always bounded from below. Because functional (1) is quadratic, in the physically relevant cases the pre-stress terms $\sigma_l (u_{xx} + u_{yy})$ and $\sigma_{zz} u_{zz}$ do not influence its stability. Thus, we can analyze the stability of the functional with zero pre-stress terms, treating the energy as the quadratic form in the variables $u_{ij}$. Actually, it is more convenient to consider $u_{xx} + u_{yy}$ and $u_{xx} - u_{yy}$ as independent variables rather than $u_{xx}$ and $u_{yy}$ separately. Apparently, $\lambda_4$ and $\lambda_5$ must be non-negative, because both related terms $(u_{xx} + u_{yy})^2 - 4(u_{xx} u_{yy} - u_{xy}^2) = (u_{xx} - u_{yy})^2 + 4u_{xy}^2$ and $u_{xz}^2 + u_{yz}^2$ are non-negative. The remaining part (terms with $\lambda_1$, $\lambda_2$, and $\lambda_3$ coefficients) corresponding to $u_{xx} + u_{yy}$ and $u_{zz}$ represents a quadratic form which must be positive semi-definite. Finally, the stability conditions are:

$$\begin{cases} \lambda_1 \geq 0, \\ \lambda_2 \geq 0, \\ \lambda_1 \lambda_2 - \lambda_3^2 \geq 0, \\ \lambda_4 \geq 0, \\ \lambda_5 \geq 0. \end{cases} \qquad (2)$$

C. Small deformation assumption

Following the notations from [47], we denote by $\nabla_i$ the covariant derivative operator with $i = 1$ or $i = 2$ corresponding to x or y, respectively. This operator equals to simple partial derivatives when it acts on scalars and vectors. Besides, the following notations are used: $\mathbf{e}_i = \nabla_i \mathbf{X}$, $\mathbf{e}'_i = \nabla_i \mathbf{X}'$, $g_{ij} = \mathbf{e}_i \cdot \mathbf{e}_j$, $\mathbf{e}^m = g^{mk} \mathbf{e}_k$, $\mathbf{T} = T^k \mathbf{e}_k = T_m \mathbf{e}^m$ with $\mathbf{e}_i$, $\mathbf{e}'_i$ being the surface basis vectors; $g_{ij}$, $g^{mk}$ — the metric tensor and its inverse; $T^k, T_m$ — the components of the tilt vector. In addition, the following equations hold: $\nabla_i \mathbf{N} = K_i^k \mathbf{e}_k$, $\nabla_i \mathbf{e}_k = -K_{ik}\mathbf{N}$, where $K_{ik}$ is the curvature tensor [52]. Following the definitions given by HK and TD, we refer to $\tilde{K}_{ik} \equiv K_{ik} + \nabla_i T_k$ as the effective curvature tensor. In terms of basis vectors and vector $\mathbf{e}'_3 = \dfrac{\partial \mathbf{X}'}{\partial z}$, the strain tensor obtains the following form:

$$\mathbf{U} = \frac{1}{2}\begin{pmatrix} \mathbf{e}'^2_1 - 1 & \mathbf{e}'_1 \cdot \mathbf{e}'_2 & \mathbf{e}'_1 \cdot \mathbf{e}'_3 \\ \mathbf{e}'_1 \cdot \mathbf{e}'_2 & \mathbf{e}'^2_2 - 1 & \mathbf{e}'_2 \cdot \mathbf{e}'_3 \\ \mathbf{e}'_1 \cdot \mathbf{e}'_3 & \mathbf{e}'_2 \cdot \mathbf{e}'_3 & \mathbf{e}'^2_3 - 1 \end{pmatrix}. \tag{3}$$

The deformations of the monolayer are assumed to be small such that the components $u_{ij}$ of the strain tensor $\mathbf{U}$ satisfy the condition $|u_{ij}| \ll 1$. Consider, for example, $u_{xx}$ at points with $z = 0$. Using the relation $2u_{xx}(z=0) = \mathbf{e}_1^2 - 1 = g_{11} - 1$, we get that $|g_{11} - 1| \ll 1$. Similarly, the consideration of $u_{xy}$ and $u_{yy}$ leads to conclusions that $|g_{12}|, |g_{22} - 1| \ll 1$. Because $2u_{zz} = \left(\dfrac{\partial \zeta}{\partial z}\right)^2 - 1$, we have $\left|\left(\dfrac{\partial \zeta}{\partial z}\right)^2 - 1\right| \ll 1$, i.e. $\left|\dfrac{\partial \zeta}{\partial z}\right| \sim 1$. Writing $u_{xz}$ at points with $z = 0$, we get $2u_{xz} = \mathbf{e}_1 \cdot \dfrac{\partial \zeta}{\partial z} \mathbf{n} = \dfrac{T_1}{\sqrt{1+\mathbf{T}^2}} \dfrac{\partial \zeta}{\partial z}$; therefore, $|T_1| \ll 1$; and, similarly, $|T_2| \ll 1$. Hence, from the assumption of $|u_{ij}| \ll 1$, the following conditions are satisfied:

$$|g_{11} - 1|, |g_{12}|, |g_{22} - 1|, |T_1|, |T_2| \ll 1. \tag{4}$$

These terms are considered to have the first order of smallness. From the equality $T_i = T^j g_{ij}$ and conditions (4), it follows that $T_i$ and $T^i$ coincide up to the first order. Therefore, $\mathbf{T}^2 = T^i T_i$ is considered to have the second order of smallness. From the equalities $2u_{xz} \approx T_1 + \nabla_1 \zeta$ and $2u_{yz} \approx T_y + \nabla_2 \zeta$, it follows that the derivatives $\nabla_i \zeta$ are at least of the first order of smallness.

In addition, we assume that the normal vector $\mathbf{N}$ and the tilt vector $\mathbf{T}$ change slowly along the dividing surface. The characteristic lengths of their change are considered to be large in comparison with the monolayer thickness $h$, which implies:

$$\left|hK_i^j\right|, \left|h\nabla_i T^j\right| \ll 1. \tag{5}$$

In fact, the same scaling rules as (5) were used in HK work [27]. In general, the spatial change of any variable is assumed to be small in this sense. For example:

$$\left| h\nabla_m hK_i^j \right| = \left| h^2 \nabla_m K_i^j \right| \ll 1. \tag{6}$$

To summarize the scaling rules, we treat the order of smallness as a number of fields occurring in the term regardless of the order of derivatives. For example, $\left|\mathbf{T}\right|$, $hK_i^j$ and $h^2\nabla_m K_i^j$ are considered as the first order of smallness terms, while $\mathbf{T}^2$, $\left(hK_i^j\right)^2$ and $h^4\left(\nabla_m K_i^j\right)^2$ — as the second order. The order of smallness of the squared effective curvature gradient is discussed further in detail in Sec. IV. In this theory of elasticity, we write the energy functional, accounting for terms up to the second order of smallness.

### D. Pre-stress terms

The pre-stress terms worth more detailed discussion. It is important to note that we are aimed at the deriving of the quadratic and stable energy functional, i.e. the monolayer energy should be bounded from below. The pre-stress terms require accurate consideration to avoid instabilities: the pre-stress itself should be small in some sense. As the pre-stress has the physical dimension (energy/length³), it cannot be directly compared with dimensionless components of the strain tensor. Therefore, the comparison with other elastic moduli is more appropriate. Analyzing the stability, one should try to find the minimum of the energy; if the minimum exists, the functional is stable. Since, by its definition, the pre-stress is multiplied by a linear deformation in the initial expression for the energy (1), while the remaining part of the energy functional is quadratic in deformations, the optimal deformation corresponding to the energy minimum should be proportional to the pre-stress. This leads to the conclusion that the pre-stress should be considered small as well as deformations.

As long as the pre-stress term is linear on deformations, it does not affect the stability. However, the polynomial parameterization of the original deformation fields can lead to the instability, unless the pre-stress itself is considered small. This can be illustrated by the simple example of elastic spring subjected to an external force. The energy of the spring of stiffness $k$ which is subjected to the external force $f = kL\varepsilon_0$ can be written as:

$$W = \frac{k}{2}x^2 - f_0 x \equiv \frac{kL^2}{2}\varepsilon^2 - \sigma_0 \varepsilon, \tag{7}$$

where $L$ is the undeformed length of the spring; $x$ is its displacement (change of the spring length); $\varepsilon = \frac{x}{L}$ and $\varepsilon_0 = \frac{f}{kL}$ are dimensionless deformations with $\varepsilon_0$ being the optimal deformation that minimizes the energy; $\sigma_0 = kL^2\varepsilon_0$ is the pre-stress. The only requirement for the stability of this system is $k > 0$. However, writing a quadratic parameterization of the deformation, for instance, in the form $\varepsilon = a_1\delta + a_2\delta^2$ and keeping only the second order in $\delta$, we obtain the following expression:

$$\frac{W}{kL^2} = \frac{1}{2}\left(a_1^2 - 2a_2\varepsilon_0\right)\delta^2 - \varepsilon_0 a_1 \delta. \tag{8}$$

The coefficient next to $\delta^2$ lacks positive definiteness, and there is the parameter space, which violates the energy stability, although originally it was stable at any deformations and pre-stress values. The reason for the stability loss is that at the high pre-stress the spring's equilibrium deformation $\varepsilon_0$ is far from being small. Although one usually assumes $x$ to be not large, the stability condition demands the energy to be bounded from below at *any* deformation, no matter how large is it. Meanwhile, the neglecting of the higher than the second order in $\delta$ in (8) was made under the assumption of the smallness of $\delta$. It is necessary to keep higher orders in $\delta$ for the energy functional to be stable at any values of parameters (considering $k > 0$). However, including higher orders makes the model nonlinear. Therefore, the equilibrium deformation $\varepsilon_0$ should be considered as the parameter of the first order of smallness ($|\varepsilon_0| \ll 1$); the term $a_2\varepsilon_0\delta^2$ in (8) thus has the third order of smallness, and, consequently, should be neglected. In this case the expression for the spring energy yields $\frac{W}{kL^2} = \frac{1}{2}a_1^2\delta^2 - \varepsilon_0 a_1 \delta$. This expression is both quadratic and stable at any values of parameters, even for an arbitrary large pre-stress $\sigma_0$ and deformation $\delta$. Although the neglect of the second-order terms next to the pre-stress reduces local quantitative accuracy, it preserves qualitative system properties, which we consider to be more important for the model. Moreover, keeping such second-order terms, one transforms the pre-stress to some kind of elastic modulus, which looks unphysical. Thus, we consider only the first order in deformation terms next to $\sigma_l$ and $\sigma_{zz}$.

## III. MONOLAYER AND BILAYER ENERGY

It this section we derive the lipid monolayer and bilayer energy using the basic notations and assumptions given in Sec. II.

### A. Incompressibility of the monolayer

Lipid bilayers have a very large volume compressibility modulus, which is approximately the same as that of water [53,54]. Furthermore, a recent study shows that lipid monolayers possess a local volume incompressibility throughout their thickness [55]. Therefore, as in Refs. [27,47], we use the assumption of the local volume incompressibility of the monolayer. Mathematically, local incompressibility means that $\left|\det(\nabla \mathbf{X}')\right| = 1$ or that $\left|\left[\frac{\partial \mathbf{X}'}{\partial x} \times \frac{\partial \mathbf{X}'}{\partial y}\right] \cdot \frac{\partial \mathbf{X}'}{\partial z}\right| = 1$. This condition allows expressing $\zeta(x,y,z)$ via $z$ and the deformation fields. We first consider the case when the dividing surface is locally non-stretchable. Formally, this can be expressed in the following form: $g = \det(g_{ij}) = 1$. In Appendix A, we give the expressions for $\frac{\partial \mathbf{X}'}{\partial x} \equiv \mathbf{e}'_1$, $\frac{\partial \mathbf{X}'}{\partial y} \equiv \mathbf{e}'_2$ and for the corresponding vector product (see (A9)). Substituting (A9) to the incompressibility condition and solving for $\zeta(z)$, we get:

$$\zeta(z) = \left(1 + \frac{1}{2}\mathbf{T}^2\right)z - \frac{1}{2}z^2 \tilde{K} + \frac{z^3}{3}\tilde{K}^2 - \frac{z^3}{3}\tilde{K}_G, \tag{9}$$

where $\tilde{K} = \tilde{K}_i^{\ i}, \tilde{K}_G = \frac{1}{2}\varepsilon^{ij}\varepsilon_{km}\tilde{K}_i^{\ k}\tilde{K}_j^{\ m}$ ($\varepsilon^{ij} \equiv \frac{\epsilon^{ij}}{\sqrt{g}}$ is the Levi-Civita tensor) are the trace and determinant of the effective curvature tensor $\tilde{K}_{ij} \equiv K_{ij} + \nabla_i T_j$. This equation is similar to the analogous equation derived by TD in Ref. [47], except for the coefficient standing at the term of the first power of $z$, which was equal to 1 in Ref. [47].

### B. Energy terms

There are four combinations of the strain tensor components in energy functional (1): $u_{xx} + u_{yy}$, $u_{zz}$, $(u_{xx} + u_{yy})^2 - 4(u_{xx}u_{yy} - u_{xy}^2)$, and $u_{xz}^2 + u_{yz}^2$. In the following, we express them via curvature and tilt fields. The result of the consideration of the first two terms up to the first order conceptually does not differ from that of HK and TD. The thorough consideration of the transverse shear combination $\left(u_{xz}^2 + u_{yz}^2\right)$ yields the tilt term and two additional contributions: tilt-curvature coupling term and squared effective curvature gradient term. In fact, with the help of (3), we can write that up to the required order:

$$4(u_{xz}^2 + u_{yz}^2) = (\mathbf{T} + \nabla \zeta)^2. \tag{10}$$

This relation differs from those obtained for $(u_{xz}^2 + u_{yz}^2)$ both by HK in [27] and by TD [47]. HK omitted the term $\nabla \zeta$. TD in Ref. [56] referred to Ref. [57] in order to explain why $u_{xz}^2 + u_{yz}^2$ equals $\frac{1}{4}\mathbf{T}^2$. However, Reddy in Ref. [58] wrote that one of the assumptions was that "the transverse normals do not experience elongation (i.e., they are inextensible)", which in our notations implies $\zeta(z) = z$. However, from (9) it follows that the function $\zeta(z)$ is more complex, and hence one may not omit the gradient of $\zeta$ on the right-hand side of (10). This fact has already been indicated in a revised version of TD's theory [48].

The right-hand side of Eq. (10) has a simple and illustrative geometrical interpretation as a square of a local tilt inside the monolayer. Recall that tilt field $\mathbf{T}$ is defined only on the dividing surface: $\mathbf{T} = \dfrac{\mathbf{n}}{\mathbf{n} \cdot \mathbf{N}} - \mathbf{N}$, where $\mathbf{n}$ is the director field, and $\mathbf{N}$ is the unit normal to the dividing surface. However, it is possible to extend this definition to the bulk of the monolayer, introducing local tilt $\mathbf{T}(z)$. To accomplish this, we consider planes which are parallel to the dividing surface in the reference configuration, when the monolayer is flat. Each of these planes deforms to surface $\mathbf{X}'(x,y,z)$ where $z$ is a fixed distance from the dividing surface to the chosen plane in the reference configuration. Now, if we denote the unit normal to surface $\mathbf{X}'(x,y,z)$ by $\mathbf{N}(z)$, we can introduce local tilt $\mathbf{T}(z)$: $\mathbf{T}(z) \equiv \dfrac{\mathbf{n}}{\mathbf{n} \cdot \mathbf{N}(z)} - \mathbf{N}(z)$. It follows that up to the quadratic order that (see Appendix B):

$$4(u_{xz}^2 + u_{yz}^2) = \mathbf{T}(z)^2. \tag{11}$$

In other words, the tilt field can be different at various $z$. For example, $\mathbf{T}(0)$ might be equal to zero, while $\mathbf{T}(z \neq 0) \neq 0$. Such a situation is demonstrated in Fig. 1(d), where the local tilt arising at $z = h$ (where $h$ is the hydrophobic thickness of the monolayer) is depicted.

Now, we consider the term $\lambda_4\left((u_{xx}+u_{yy})^2 - 4(u_{xx}u_{yy} - u_{xy}^2)\right)$. The modulus $\lambda_4$ corresponds to the lateral shear [51]. Because of the local lateral fluidity of the monolayer, we assume that $\lambda_4 = 0$, which leads to a twist modulus being equal to zero (see Appendix C). This assumption is conventional for treating fluids (for the review see [59]). The weaker assumption of the global fluidity, $\int \lambda_4(z)\,dz = 0$, could be made. However, the stability condition $\lambda_4 \geq 0$, obtained in (2), implies the equivalence of both local and global fluidity assumptions. In addition, we stress that the free energy (1) refers to an equilibrium state: the lipid director and tilt should be considered as time-averaged quantities, whereas the moduli in this free energy reflect the energy cost of a variation of these time-averaged quantities. The fluidity conditions and equality $\lambda_4 = 0$ should also be interpreted in this context, albeit transverse motions of individual lipids might be hindered at small time scales.

Note that HK made slightly different assumptions about the monolayer fluidity. Writing the expression for $(u_{xx}+u_{yy})$ in terms of the combination $\tilde{u}^2 = \left((u_{xx}+u_{yy})^2 - 4(u_{xx}u_{yy} - u_{xy}^2)\right)$ and the lateral strain $\varepsilon = \varepsilon(x,y,z) \equiv \|\mathbf{e}_1' \times \mathbf{e}_2'\|$, HK obtained that the modulus corresponding to $\tilde{u}^2$ equals $\sigma_l + \dfrac{\lambda_4}{2}$. Then they considered both cases corresponding to two fluidity assumptions: strong $\sigma_l + \dfrac{\lambda_4}{2} = 0$ and weak $\int \left(\sigma_l + \dfrac{\lambda_4}{2}\right) dz = 0$. However, if $\tilde{u}^2$ is considered as an independent deformation mode, one should require the corresponding modulus to be non-negative. Therefore, strong and weak assumptions are again equivalent. A more detailed discussion of the assumptions made by HK is provided in Appendix C.

### C. Final expression for bilayer and monolayer energy

Substituting relations (3) and (10) into Eq. (1), we arrive at the following expression for the energy density of a single lipid monolayer:

$$F = \sigma_0(z)z\tilde{K} + \frac{1}{2}E(z)z^2\tilde{K}^2 + \frac{1}{2}\lambda_T(z)\mathbf{T}(z)^2$$
$$= \sigma_0(z)z\tilde{K} + \frac{1}{2}E(z)z^2\tilde{K}^2 + \frac{1}{2}\lambda_T(z)\left[\mathbf{T} - \frac{1}{2}z^2\boldsymbol{\nabla}\tilde{K}\right]^2, \tag{12}$$

where:

$$\sigma_0(z) = \sigma_l(z) - \sigma_{zz}(z), \tag{13a}$$

$$E(z) = 4\lambda_1(z) + \lambda_2(z) - 4\lambda_3(z), \tag{13b}$$

$$\lambda_T(z) = 2\lambda_5(z). \tag{13c}$$

Integrating the energy density with respect to $z$ over the monolayer thickness from the lower to upper boundary of the monolayer, we get the expression for the surface energy density of a single lipid monolayer:

$$e_{2d}^{mono} = \frac{1}{2}k_m(\tilde{K} - K_{0,m})^2 - \frac{1}{2}k_m K_{0,m}^2 + \\ + \frac{1}{2}k_t \mathbf{T}^2 + k_c \mathbf{T} \cdot \boldsymbol{\nabla}\tilde{K} + \frac{k_{gr}}{2}(\boldsymbol{\nabla}\tilde{K})^2, \tag{14}$$

where:

$$k_m = \int dz\, E(z) z^2, \tag{15a}$$

$$-k_m K_{0,m} = \int dz\, \sigma_0(z)\, z, \tag{15b}$$

$$k_t = \int dz\, \lambda_T(z), \tag{15c}$$

$$k_c = -\frac{1}{2}\int dz\, \lambda_T(z)\, z^2, \tag{15d}$$

$$k_{gr} = \frac{1}{4}\int dz\, \lambda_T(z)\, z^4. \tag{15e}$$

In order to obtain the total elastic energy of the monolayer, the integration of expression (14) over the plane $z = 0$ can be replaced by the integration over the deformed dividing surface because of the assumption of the dividing surface local inextensibility, which implies $\sqrt{g} = 1$. Therefore, the energy density (14) is invariant under reparametrizations of the dividing surface. In comparison with the results obtained by TD [47,48], the term $\frac{k_{gr}}{2}(\boldsymbol{\nabla}\tilde{K})^2$ in expression (14) is new. Noting that the director divergence $\boldsymbol{\nabla} \cdot \mathbf{n} \equiv \mathbf{e}^i \nabla_i \cdot \mathbf{n}$ up to the first order equals to $\tilde{K}$, we can rewrite (14) as:

$$e_{2d}^{mono} = \frac{1}{2}k_m(\boldsymbol{\nabla}\cdot\mathbf{n} - K_{0,m})^2 - \frac{1}{2}k_m K_{0,m}^2 + \\ + \frac{1}{2}k_t \mathbf{T}^2 + k_c \mathbf{T}\cdot(\boldsymbol{\nabla}\boldsymbol{\nabla}\cdot\mathbf{n}) + \frac{k_{gr}}{2}(\boldsymbol{\nabla}\boldsymbol{\nabla}\cdot\mathbf{n})^2, \tag{16}$$

where $\boldsymbol{\nabla}\boldsymbol{\nabla}\cdot\mathbf{n} \equiv \boldsymbol{\nabla}(\boldsymbol{\nabla}\cdot\mathbf{n})$ is the gradient of the director divergence.

To derive Eq. (16) we assumed that lipid heads behave like lipid tails during tilting and therefore the direction of their director coincides with that of lipid tails. It is, however, possible that they have their own director independent of the direction of lipid tails. Then, additional terms corresponding to lipid heads director should be taken into account in Eq. (16). If the tilt modulus of lipid heads is high, then lipid heads stay approximately normal to the dividing surface and the corresponding director divergence may be replaced with curvature, giving rise to bare bending contributions sometimes considered in the literature [60].

### D. Accounting for the lateral stretching-compression

Above we considered only the deformations with a locally constant area of the dividing surface. Here, we additionally address the lateral stretching-compression deformation. We attribute monolayer deformations to an arbitrary surface inside the monolayer and derive the expressions for the elastic moduli corresponding to this new surface. The chosen surface is assumed to be parallel to the dividing surface in the reference configuration when the monolayer is flat. A Cartesian coordinate system is now chosen is such a way that the $xy$-plane coincides with the new chosen surface, and we denote by $\alpha$ a local area variation of this surface. Now, vector product (A9) should be multiplied by $1+\alpha$ and therefore expression (9) for $\zeta(z)$ transforms to expression (A11). The expression for $u_{xx}+u_{yy}$ and $u_{zz}$ up to the first order can be written as:

$$u_{xx}+u_{yy} = -u_{zz} = \alpha + \zeta \tilde{K}. \tag{17}$$

Therefore, expressions (12), (14) and (16) for the energy density of a single monolayer become:

$$F = \sigma_0(z)(\alpha + z\tilde{K}) + \frac{1}{2}E(z)(\alpha + z\tilde{K})^2 + \frac{1}{2}\lambda_T(z)[\mathbf{T} - z\boldsymbol{\nabla}\alpha - \frac{1}{2}z^2\boldsymbol{\nabla}\tilde{K}]^2, \tag{18}$$

$$\begin{aligned}e_{2d}^{mono} &= \frac{1}{2}k_m(\tilde{K}-K_{0,m})^2 - \frac{1}{2}k_m K_{0,m}^2 \\ &+ \frac{1}{2}k_{t,m}\mathbf{T}^2 + k_c\mathbf{T}\cdot\boldsymbol{\nabla}\tilde{K} + \frac{k_{gr}}{2}(\boldsymbol{\nabla}\tilde{K})^2 \\ &+ \frac{1}{2}k_A(\alpha-\alpha_{0,m})^2 - \frac{1}{2}k_A\alpha_{0,m}^2 + A\alpha\tilde{K} \\ &+ B\mathbf{T}\cdot\boldsymbol{\nabla}\alpha - k_c(\boldsymbol{\nabla}\alpha)^2 + C\boldsymbol{\nabla}\alpha\cdot(\boldsymbol{\nabla}\tilde{K}),\end{aligned} \tag{19}$$

$$\begin{aligned}e_{2d}^{mono} &= \frac{1}{2}k_m(\boldsymbol{\nabla}\cdot\mathbf{n}-K_{0,m})^2 - \frac{1}{2}k_m K_{0,m}^2 \\ &+ \frac{1}{2}k_{t,m}\mathbf{T}^2 + k_c\mathbf{T}\cdot(\boldsymbol{\nabla}\boldsymbol{\nabla}\cdot\mathbf{n}) + \frac{k_{gr}}{2}(\boldsymbol{\nabla}\boldsymbol{\nabla}\cdot\mathbf{n})^2 \\ &+ \frac{1}{2}k_A(\alpha-\alpha_{0,m})^2 - \frac{1}{2}k_A\alpha_{0,m}^2 + A\alpha\boldsymbol{\nabla}\cdot\mathbf{n} \\ &+ B\mathbf{T}\cdot\boldsymbol{\nabla}\alpha - k_c(\boldsymbol{\nabla}\alpha)^2 + C\boldsymbol{\nabla}\alpha\cdot(\boldsymbol{\nabla}\boldsymbol{\nabla}\cdot\mathbf{n}),\end{aligned} \tag{20}$$

where the new moduli are given by:

$$k_A = \int dz\, E(z), \qquad (21a)$$

$$-k_A \alpha_{0,m} = \int dz\, \sigma_0(z), \qquad (21b)$$

$$A = \int dz\, E(z)\, z, \qquad (21c)$$

$$B = -\int dz\, \lambda_T(z)\, z, \qquad (21d)$$

$$C = \frac{1}{2} \int dz\, \lambda_T(z) z^3, \qquad (21e)$$

and the limits of the integration depend on the position of the new reference plane. To obtain the total elastic energy of the monolayer, the surface integral of expressions (19) and (20) should be performed over the plane $z = 0$. However, all terms in (19) and (20) are quadratic, and therefore the total energy of the monolayer can be obtained via the integration over the deformed surface without losing any contributions up to the second order. Thus, the energy becomes invariant under reparametrizations of the chosen surface. In view of conditions (2), expression (18), and consequently the energy functionals (19) and (20), are bounded from below. Note also that the coupling modulus $k_c$ coincides up to a sign with that of the stretching gradient modulus. Such a stretching gradient term was also previously derived in Ref. [61] within the assumptions similar to those made by HK [27] and was indicated as crucial to explain some experiments [62]. The new contributions to the elastic energy of lipid membranes are illustrated in Fig. 1.

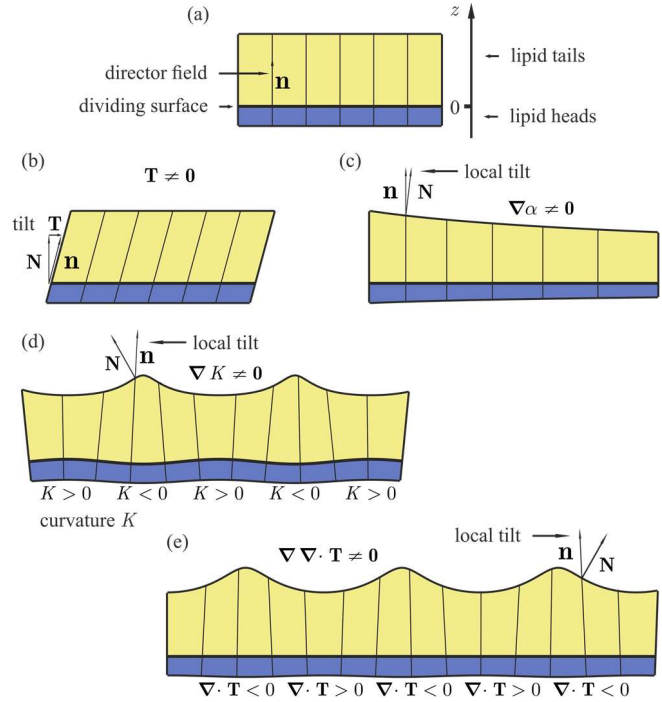

FIG. 1. (Color online) The illustration of the new contributions to the elastic energy of lipid membranes. For the sake of simplicity, all deformations are presented in the one-dimensional case. All shapes are exactly calculated under the incompressibility constraint. a) The undeformed lipid

monolayer in the reference configuration. The hydrophobic thickness of the monolayer is chosen to be 1.5 nm; the thickness of the lipid heads region is set 0.4 nm for illustrative purposes. b) The pure tilt deformation at an angle of 20°. c) The constant stretching gradient deformation with $\frac{d\alpha}{dx} = 0.1$ nm$^{-1}$. d) The pure bending deformation of the dividing surface (no stretching and zero tilt with respect to the dividing surface) to a sine wave $0.05\sin(2x)$. e) The deformation of zero curvature and varying tilt divergence: $T_x = 0.05\sin(2x)$, where $T_x$ is the $x$-component of the tilt field. Panels (c), (d) and (e) show that a local tilt, which is responsible for the new contributions, can emerge inside the lipid monolayer even if there is no tilt with respect to the dividing surface or it is small as in panel (e). Such a local tilt emerges due to thickness variations which result in transverse shear deformations inside a lipid monolayer. For illustrative purposes, the local tilt in panels (c), (d) and (e) is shown for surfaces which in the reference configuration correspond to planes parallel to the dividing surface at a distance equal to the hydrophobic thickness of the monolayer.

## IV. DISCUSSION

In this paper, we revisited the elastic energy functionals derived by HK and TD [27,47]. We demonstrated that both functionals should be appended by the additional term proportional to the squared gradient of the director divergence. In addition, the influence of the pre-stress on the stability of the energy functional was considered. It was shown that in order to ensure the stability the effective Gaussian curvature term should be omitted.

Under the assumption of the inextensibility of the dividing surface, we derived that the energy functional obtained by HK should be appended by two additional terms: $\mathbf{T} \cdot (\boldsymbol{\nabla} \tilde{K}) \approx \mathbf{T} \cdot (\boldsymbol{\nabla} \boldsymbol{\nabla} \cdot \mathbf{n})$ and $(\boldsymbol{\nabla} \tilde{K})^2 \approx (\boldsymbol{\nabla} \boldsymbol{\nabla} \cdot \mathbf{n})^2$; the energy functional obtained by TD — by one additional term $(\boldsymbol{\nabla} \tilde{K})^2 \approx (\boldsymbol{\nabla} \boldsymbol{\nabla} \cdot \mathbf{n})^2$. Below, the key points of the present work are discussed in detail: the order of smallness of the $(\boldsymbol{\nabla} \tilde{K})^2$ term, the neglection of the effective Gaussian curvature term, the physical consequences of the new energy terms and their influence on the director fluctuations, and the use of the incompressibility constraint in the energy functional.

## A. $(\nabla \tilde{K})^2$ term

There is no term proportional to $(\nabla \tilde{K})^2$ in the energy functionals obtained by HK [27] and TD [56]. In our derivation of the monolayer energy (14) the term $\frac{k_{gr}}{2}(\nabla \tilde{K})^2$ comes from the integral $\frac{1}{8}\int dz\, \lambda_T(z) z^4 (\nabla \tilde{K})^2$. Applying the scaling rules introduced in Eq. (6), we conclude that the combination $h^4 (\nabla \tilde{K})^2$ is of the second order of smallness. Because the integration is performed over the monolayer thickness, the term $\frac{k_{gr}}{2}(\nabla \tilde{K})^2$ turns out to be of the same order as the terms $\frac{1}{2}\int dz\, \lambda_T(z) \mathbf{T}^2$ and $-\frac{1}{2}\int dz\, \lambda_T(z) z^2\, \mathbf{T}\cdot\nabla\tilde{K}$ corresponding to the tilt and tilt-curvature coupling, respectively.

Another argument to keep the term $(\nabla \tilde{K})^2$ in the energy functional is based on the stability considerations. The energy density (12) is stable in terms of boundedness from below. Indeed, in view of stability conditions (2), we have that $\lambda_T(z) \geq 0$ and $E(z) = \frac{1}{\lambda_1(z)}\left\{\left[\lambda_1(z)\lambda_2(z) - \lambda_3^2(z)\right] + \left[2\lambda_1(z) - \lambda_3(z)\right]^2\right\} \geq 0$. Thus, the second and third terms in the energy density (12) are non-negative. As for the first term, it does not violate boundedness of the functional, being the linear term in $z\tilde{K}$ in the presence of the non-negative second-order term $(z\tilde{K})^2$. Omitting the term proportional to $(\nabla \tilde{K})^2$ ruins the boundness of the third term, leading to the instability of the whole functional. This explains the instability of the energy functional obtained by TD as discussed in Ref. [47]. Another argument is that the term $\left[\mathbf{T} - \frac{1}{2}z^2 \nabla \tilde{K}\right]^2$ in (12) originally comes from the expression $u_{xz}^2 + u_{yz}^2$, which is always non-negative, while neglecting $(\nabla \tilde{K})^2$ breaks this property.

The term $(\nabla \tilde{K})^2$ includes spatial derivatives of the second order. They appear because of the deformation parameterization in the form of $\mathbf{X}'(x,y,z) = \mathbf{X}(x,y) + \zeta(x,y,z)\mathbf{n}(x,y)$ and the incompressibility assumption, which leads to expression (9) for $\zeta$ and for its gradient:

$\nabla \zeta = -\frac{1}{2} z^2 \nabla \tilde{K} \approx -\frac{1}{2} z^2 \nabla \nabla \cdot \mathbf{n}$. Therefore, although there are only the first-order derivatives in initial functional (1), the derivatives of the second order appear in the final relation. Actually, there are theories of elasticity that consider the energy as an explicit function of the strain gradient [63,64] and even the strain gradients of higher orders [65,66]. In our case, after the incorporation of the strain gradients up to the *n*-th order in energy functional (1), the order of derivatives in the final answer will be $n + 1$. Nevertheless, this number is always restricted, and therefore scaling rules (5), (6) cause no formal mathematical problems.

Terms $\mathbf{T}^2$, $\mathbf{T} \cdot \nabla \tilde{K}$ and $(\nabla \tilde{K})^2$ originate from transverse shear deformation. These terms were not present in previous theories since, as was already mentioned in the introduction, these theories did not address the transverse isotropy of lipid monolayers. There are no such terms in Ref. [28] because the classical Frank theory, which was originally derived for liquid crystals from other symmetry arguments, is used there. As for the opposing forces model [16,29,30], it considers only three energy contributions: the repulsion of lipid heads, surface tension and lipid stretching. The first two terms are related to $u_{xx} + u_{yy}$ in Eq. (1), whereas the third term corresponds to $u_{zz}$. We therefore see that the opposing forces model lacks transverse shear deformations $u_{xz}$ and $u_{yz}$. The tilt term, which is nevertheless present in the opposing forces model, corresponds to keeping the quadratic order in the pre-stress in Eq. (1) because $\mathbf{T}^2$ is present in $u_{zz}$. We, however, keep only first-order terms due to stability arguments (see also Sec. IV.C), although $\mathbf{T}^2$ in the pre-stress may be regarded as a correction to the tilt modulus [67].

### B. Estimation of the new moduli $k_c$, $k_{gr}$, B and C

We can roughly estimate the new moduli $k_c$ and $k_{gr}$, assuming that the main contribution to the tilt modulus $k_t = \int dz\, \lambda_T(z)$ comes from the hydrophobic region of the monolayer and that $\lambda_T(z)$ is approximately constant in this region. If the length of hydrocarbon chains in the reference undeformed state of the monolayer is $l$, then $\lambda_T(z) \approx \frac{k_t}{l}$, and from Eqs. (15e), (15d), (21d) and (21e) we obtain $k_c \approx -\frac{k_t l^2}{6}$, $k_{gr} \approx \frac{k_t l^4}{20}$, $B \approx -\frac{k_t l}{2}$, $C \approx \frac{k_t l^3}{8}$. Using HK's estimation of the tilt modulus $k_t \approx 50$ mN/m $\approx 12\, k_B T/\text{nm}^2$ ($k_B$ is Boltzmann constant, $T = 300$ K) [27] and common value of the hydrocarbon chain length $l \approx 1.5$ nm, we get $k_c \approx -5\ k_BT$, $k_{gr} \approx 3\ k_BT\cdot\text{nm}^2$, $B \approx -9\ k_BT/\text{nm}$, $C \approx 5\ k_BT\cdot\text{nm}$.

## C. Neglecting the effective Gaussian curvature

HK [27] and TD [47] keep the effective Gaussian curvature term $\tilde{K}_G = \frac{1}{2}\varepsilon^{ij}\varepsilon_{km}\tilde{K}_i^{\ k}\tilde{K}_j^{\ m}$, which formally has the second order of smallness and originates from the pre-stress term $\sigma_0(z)\varepsilon(z)$, where $\varepsilon(z) = \varepsilon(\zeta(z)) = \zeta\tilde{K} + \zeta^2\tilde{K}_G + \mathbf{T}\cdot\nabla\zeta + \frac{1}{2}(\nabla\zeta)^2$ is the lateral strain, the expression for which is obtained in (A10). The Gaussian curvature is widely used in the elasticity theory of lipid membranes; its elastic modulus was determined both experimentally and from molecular dynamics simulations [68–70]. However, we argue that within the framework of the simplest classical theory of elasticity presented this term is mathematically and physically doubtful, as the accounting for this term leads to an excessive accuracy.

Firstly, the effective Gaussian curvature term leads to the instability of the energy functional in the case of free boundary conditions for the tilt field. Underlying energy functional (1) is always bounded from below as long as (2) holds. But one violates the stability of the functional, keeping the effective Gaussian curvature (which has the second order of smallness) in the pre-stress part. This can be illustrated by the following simple example. Consider the flat monolayer with the dividing surface occupying the square region $(0,a)\times(0,a)$ in $xy$-plane with $a > 0$. We choose a simple parameterization using a Cartesian coordinate system and assume the projections of the tilt onto $x$ and $y$ axes to be the functions of only $y$ and $x$, respectively, i.e. $T_x = T_x(y)$ and $T_y = T_y(x)$. This implies that the mean curvature is zero and leads to the simplified expression for the Gaussian curvature: $\tilde{K}_G = -\frac{\partial T_x}{\partial y}\frac{\partial T_y}{\partial x}$. Now, the total energy density takes the form:

$$e_{2d}^{mono} = -\bar{k}_m\frac{\partial T_x}{\partial y}\frac{\partial T_y}{\partial x} + \frac{1}{2}k_t(T_x^2 + T_y^2), \qquad (22)$$

where $\bar{k}_m$ is the effective Gaussian curvature modulus. The full energy is the integral over the $(0,a)\times(0,a)$ region. After integration by parts of the first term, the energy is given by:

$$-\bar{k}_m(T_x(a) - T_x(0))(T_y(a) - T_y(0)) + \frac{1}{2}k_t\iint(T_x^2 + T_y^2)\,dxdy. \qquad (23)$$

Assume now that $T_x^2$ and $T_y^2$ approach the $\delta$-functions with its peak on the boundary of the chosen region. For example:

$$T_x^2 = \frac{h}{|\beta|\sqrt{\pi}} e^{-\left(\frac{y}{\beta}\right)^2}, \quad T_y^2 = \frac{h}{|\beta|\sqrt{\pi}} e^{-\left(\frac{x}{\beta}\right)^2}, \tag{24}$$

where the multiplier in the form of the monolayer thickness $h$ was introduced in order to keep the tilt dimensionless. We assume that projections $T_x$ and $T_y$ are positive. While $\iint (T_x^2 + T_y^2) dx dy \to h \cdot a$, $(T_x(a) - T_x(0))(T_y(a) - T_y(0)) \to +\infty$ as $\beta \to 0$. On the other hand, if we choose the tilt to approach the $\delta$-function on the other side of the region, for instance,

$$T_x^2 = \frac{h}{|\beta|\sqrt{\pi}} e^{-\left(\frac{y}{\beta}\right)^2} \quad \text{and} \quad T_y^2 = \frac{h}{|\beta|\sqrt{\pi}} e^{-\left(\frac{x-a}{\beta}\right)^2}, \quad \text{then} \quad (T_x(a) - T_x(0))(T_y(a) - T_y(0)) \to -\infty \text{ as}$$

$\beta \to 0$. Therefore, the functional is unstable, because it lacks the lower bound regardless of the sign of $\overline{k}_m$ in the case of the free boundary conditions for the tilt field.

There are no such stability problems in the classical Helfrich functional [7]: $w = \frac{1}{2}k(K - c_0)^2 + \overline{k}K_G$ with $k$ and $\overline{k}$ being the bending and Gaussian curvature moduli, respectively, because the mean and Gaussian curvatures can be written via the sum and product of the principal curvatures $c_1$, $c_2$, and one needs only to require the positive semidefiniteness of a quadratic form $\frac{1}{2}k(c_1 + c_2)^2 + \overline{k}c_1 c_2 > 0 \Leftrightarrow -2k \leq \overline{k} \leq 0 \ \& \ k \geq 0$ for the stability. However, such decomposition cannot be made when the effective mean and Gaussian curvatures are considered because effective curvature tensor $\tilde{K}_{ij}$ is not symmetric. Therefore, the replacement of mean $K$ and Gaussian $K_G$ curvatures by the effective ones ($\tilde{K}$ and $\tilde{K}_G$) in the Helfrich functional leads to the loss of boundedness, even if the functional is appended by the tilt squared. However, the addition of the twist term $\frac{k_{tw}}{2}(\nabla \times \mathbf{T})^2 \equiv \frac{k_{tw}}{2}\left(\varepsilon^{ij}\nabla_i T_j\right)^2 = \frac{k_{tw}}{2}\left(\varepsilon^{ij}\tilde{K}_{ij}\right)^2$, where $\varepsilon^{ij}$ is the Levi-Civita tensor and $k_{tw}$ is the twist modulus, stabilizes the functional, which in this case has the form:

$$\tilde{w} = \frac{1}{2}k\left(\tilde{K} - c_0\right)^2 + \overline{k}\tilde{K}_G + \frac{k_{tw}}{2}(\nabla \times \mathbf{T})^2 + \frac{1}{2}k_t\mathbf{T}^2. \tag{25}$$

Expression (25) is written in the form suggested by Helfrich [7], i.e. without the tilt-curvature coupling and curvature gradient. The stability of this functional is convenient to analyze in a local

Cartesian coordinate system with its origin in a given point of the monolayer reference surface, where the functional can be written as:

$$\tilde{w} = \frac{1}{2}k\left(\tilde{K}_{11} + \tilde{K}_{22} - c_0\right)^2 + \bar{k}\left(\tilde{K}_{11}\tilde{K}_{22} - \tilde{K}_{12}\tilde{K}_{21}\right) + \frac{k_{tw}}{2}\left(\tilde{K}_{12} - \tilde{K}_{21}\right)^2 + \frac{1}{2}k_t\mathbf{T}^2. \qquad (26)$$

Treating it as a quadratic form in $\tilde{K}_{ij}$ plus the tilt squared, amended by the linear terms, one obtains the following stability conditions: $-2k \leq \bar{k} \leq 0 \ \& \ -2k_{tw} \leq \bar{k} \leq 0 \ \& \ k, k_{tw}, k_t \geq 0$. However, because of the lateral fluidity of the monolayer and stability conditions (2) requiring $\lambda_4 \geq 0$, the lateral shear modulus $\lambda_4$ is assumed to be equal to zero in energy functional (1), and hence there is no twist term in the final expression for the monolayer energy (14), and therefore the effective Gaussian curvature contribution cannot be stabilized by the twist term.

Secondly, an argument against the Gaussian curvature term is that the corresponding modulus $\bar{k}_m = \int dz\,\sigma_0(z)\,z^2$ contains only the pre-stress parameter of the system, but not any elastic moduli of underlying functional (1), which looks contradictory. Both stability and modulus arguments suggest that it is excessive to keep the second order in the pre-stress terms since it leads to the emergence of the artificial elastic modulus and instability of the initially stable monolayer. Although there is no Gaussian curvature in the monolayer energy (14), it does not imply that the monolayer deformations with a zero mean curvature and non-zero Gaussian curvature cost no energy. Formally, from (14) it follows that the energy of such deformations is zero, but actually, this contribution was just neglected from the functional as formally being of a higher order of smallness. To retain both the energetic contribution of the Gaussian curvature and the stability of the monolayer, one needs to include the term proportional to $\tilde{K}_G^2$. In this case, $\bar{k}_m$ enters the expression for the spontaneous effective Gaussian curvature. However, such a contribution to the energy is of the higher order of smallness and makes the model nonlinear in terms of principal curvatures. Keeping only the first-order term in $\tilde{K}_G$ is analogous to keeping only the first-order term in $\tilde{K}$ and neglecting the second-order one, which obviously makes the monolayer unstable. A similar argument should be applied to the opposing forces model [16], where the functional should be amended by $\tilde{K}_G^2$ coming from the lipid chain's conformational free energy, as this free energy acts there as a stabilizing contribution.

In Ref. [71], the problem concerning the calculation of the Gaussian curvature modulus from simulations via the expression $\bar{k}_m = \int dz\,\sigma_0(z)\,z^2$ was stated, as this formula often yields positive values [72,73], which are beyond the stability range. In this paper, we showed that in the quadratic model both positive and negative signs of $\bar{k}_m$ lead to the instability, which can be avoided only by higher-order corrections. Thus, the positive values of $\bar{k}_m$ found in Refs. [72,73] do not represent a problem. Actually, in the same way as $\int dz\,\sigma_0(z)\,z$ is associated with the spontaneous effective curvature via the expression $-k_m K_{0,m} = \int dz\,\sigma_0(z)\,z$, , the expression $\int dz\,\sigma_0(z)\,z^2$ can be viewed as the spontaneous effective Gaussian curvature, and therefore can be of either sign. This can be demonstrated by considering a constant pure bending deformation up to the fourth order under conditions of zero tilt and inextensible dividing surface. In this case, the energy density takes the following form:

$$w_{ls}^{3d} = \sigma_0(z)\left[zK + z^2 K_G - \frac{z^2}{2}K^2\right] + \frac{1}{2}E(z)\left[zK + z^2 K_G - \frac{z^2}{2}K^2\right]^2, \tag{27}$$

with the corresponding two-dimensional contributions being:

$$\begin{aligned}w_{ls}^{2d} &= \frac{1}{2}k_m\left(K - K_{0,m}\right)^2 - \frac{1}{2}k_m K_{0,m}^2 + \frac{1}{2}\bar{k}_{2m}\left(K_G - K_{G0,m}\right)^2 - \frac{1}{2}\bar{k}_{2m}K_{G0,m}^2 \\ &+ \frac{1}{8}\bar{k}_{2m}\left(K^2 + 2K_{G0,m}\right)^2 - \frac{1}{2}\bar{k}_{2m}K_{G0,m}^2 + k_{2c}KK_G - \frac{1}{2}k_{2c}K^3 - \frac{1}{2}\bar{k}_{2m}K^2 K_G,\end{aligned} \tag{28}$$

where:

$$\bar{k}_{2m} = \int dz\,E(z)z^4, \tag{29a}$$

$$-\bar{k}_{2m}K_{G0,m} = \int dz\,\sigma_0(z)\,z^2, \tag{29b}$$

$$k_{2c} = \int dz\,E(z)z^3. \tag{29c}$$

From Eqs. (28), (28a-b), one can see that the integral $\int dz\,\sigma_0(z)\,z^2$ plays the role similar to that of $\int dz\,\sigma_0(z)\,z$, as both of them are involved in the expressions for the corresponding spontaneous curvatures.

A notion of "spontaneous warp" similar to the spontaneous Gaussian curvature was introduced by T. M. Fischer [74], where the contribution of deviatoric bending in the form $\sim\left(\left(c_1 - c_2\right)/2 - \vartheta\right)^2$, reflecting the difference between principal curvatures $c_1$ and $c_2$ is considered. This contribution comes from lateral shear stresses, which we neglect due to the lateral fluidity of lipid monolayers. In Ref. [74] it was also indicated that such stresses relax much faster than curvature stresses. The Gaussian curvature term $\sim\left(K_G - K_{G0,m}\right)^2$ (see Eq. (28)) contains fourth powers of $c_1$, $c_2$, and therefore definitely differs from deviatoric bending, and it has a different nature, coming not from lateral shear but rather from the bending deformation as a higher-order correction to mean curvature bending. $K_{G0,m}$ may be important in situations when the regions with high $|c_1 - c_2|$ are considered. For example, the presence of membrane components with nonzero $K_{G0,m}$ may lead to their segregation to the neck region during fission, in a similar way to that proposed in Refs. [75–77]. As for the fission event, it might occur as a result of the formation of pores or other topological defects [78].

### D. Physical implications

In order to demonstrate the consequences of the correction of the energy functional, we consider a simple and illustrative problem, in which tilt modes still contribute significantly. We calculate the distance dependence of the energy of the interaction of two amphipathic peptides, mediated by the elastic deformations of a bilayer. At large distances between two peptides partially incorporated into the same monolayer, the induced membrane deformations are independent, and the corresponding energies are additive. However, at smaller distances, these deformations overlap and lead to the effective lateral interaction of two peptides. In general, such an interaction occurs not only between amphipathic peptides but also between other inclusions incorporated into lipid membranes [79–83]. Recently, it was shown in Ref. [84] that the interaction of two amphipathic peptides can be accurately described via a one-dimensional approach within the framework of HK's Hamiltonian. We will use the same approach and compare the results obtained in the framework of Hamiltonian (16) with that following from the HK Hamiltonian. A shallowly incorporated amphipathic peptide is modeled as a cylinder, one side surface of which is hydrophilic and the other is hydrophobic; the axis of the cylinder is assumed to lie in the plane of the membrane. The peptide induces a tilt in the adjacent monolayer and therefore a director jump, $\Delta \mathbf{n} = \mathbf{n}_2 - \mathbf{n}_1$, at its boundaries. Let $H_u(x)$ be the shape of the dividing surface of the upper monolayer: the distance between the reference plane and the dividing surface, measured along the normal to the reference plane. We allow amphipathic peptides to rotate around their longitudinal axis, which implies the following boundary condition for the dividing surface of the adjacent monolayer: $H_u(x_0 + (D/2)) - H_u(x_0 - (D/2)) = D(n_{1x} + n_{2x})$, where $x_0$ is the coordinate of the peptide center; $n_{1x}$ and $n_{2x}$ are the director projections onto the $x$-axis at the left and right boundaries of the peptide; $D$ is the width (diameter) of the peptide, which we assume to be 1.3 nm, i.e. approximately the diameter of an $\alpha$-helix. The parameters of the bilayer are the following: bending modulus $k_m$ = 10 $k_BT$ [85], tilt modulus $k_t$ = 12 $k_BT$/nm$^2$ [27], and zero spontaneous curvatures. Hydrophobic thickness $h$, which reflects the mean ordering of the hydrocarbon chains of lipids [86], is set to 1.5 nm [85]. We also add lateral tension terms to the energy functional: $\sigma \cdot \left( \sqrt{1 + \frac{d}{dx} H_u^2} - 1 \right) \approx \frac{1}{2} \sigma \cdot \left( \frac{d}{dx} H_u \right)^2$ and $\sigma \cdot \left( \sqrt{1 + \frac{d}{dx} H_l^2} - 1 \right) \approx \frac{1}{2} \sigma \cdot \left( \frac{d}{dx} H_l \right)^2$ per unit length along the *y*-axis, with σ = 0.0025 $k_BT$/nm$^2$ [87] in each monolayer. We use the same qualitative estimation of the director jump as in Ref. [84], namely $n_{2x} - n_{1x} = \Delta n_x = -\frac{D}{\sqrt{(D/2)^2 + h^2}} \approx -0.8$. For the sake of simplicity, we neglect stretching-compression deformation mode, assuming the elastic modulus of this mode to be large in comparison with other moduli [85]. This leaves us with two unknown moduli $k_c$ and $k_{gr}$, which we vary in order to investigate their influence on the interaction of the peptides. It is important to note that $k_c$ and $k_{gr}$ cannot take arbitrary values because they must satisfy the Cauchy-Bunyakovsky inequality $k_{gr} k_t - k_c^2 \geq 0$, which follows from Eqs. (15c–e) and stability requirements (2). Keeping $k_c$ and $k_{gr}$ within this permissible condition, we solve Euler-Lagrange equations for functional (16) (see Appendix D) and find the minimum energy at various distances between two amphipathic peptides under the boundary conditions described above. Solutions of Euler-Lagrange equations for the HK Hamiltonian can be found in Ref. [39].

The results are presented in Fig. 2. Firstly, we fix $k_c = 0$ and vary $k_{gr}$ (Fig. 2(a)). As we see, the HK Hamiltonian predicts the global energy minima at a distance of about 4 nm between the peptides. This energy minimum remains almost unchanged when $k_{gr}$ is varied: increasing the values of $k_{gr}$ shifts the energy profile upwards without significantly changing its shape. Predictably, the energy profiles obtained from Hamiltonian (16) approach HK's energy profile as $k_c, k_{gr} \to 0$. Figure 2(c) shows the energy profile with theoretically estimated values of moduli ($k_c \approx -5$ $k_BT$ and $k_{gr} \approx 3$ $k_BT$·nm$^2$) compared to the profile following from the HK Hamiltonian: the depth of the energy well drops from 0.27 $k_BT$/nm in the case of HK's Hamiltonian to 0.15 $k_BT$/nm in the case of Hamiltonian (16), demonstrating an almost two-fold decrease. Next, fixing $k_{gr} = 12$ $k_BT$·nm$^2$, we vary the coupling modulus $k_c$ in order to explore how $k_c$ effects the energy profile. Recall that, according to Eq. (15d) and stability conditions (2), $k_c < 0$. In addition, the stability condition $k_{gr} k_t - k_c^2 \geq 0$ together with $k_t = 12$ $k_BT$/nm$^2$ implies that $k_c \geq -12$ $k_BT$. Varying $k_c$ within this range, we obtain the curves presented in Fig. 2(d). One can see that an increase in the absolute value of $k_c$ is accompanied by a decrease in the depth of the energy well, which eventually disappears. For example, there is no local energy minimum at $k_c = -11.5\ k_B T$. In particular, it turns out that at the global minimum in HK's case the vectors of tilt and effective curvature gradient are antiparallel at all points of the upper monolayer between the two peptides. Therefore, accounting for $k_c \mathbf{T} \cdot (\nabla \tilde{K})$ and $\frac{k_{gr}}{2}(\nabla \tilde{K})^2$ increases the energy at this point due to the fact that $k_c < 0$.

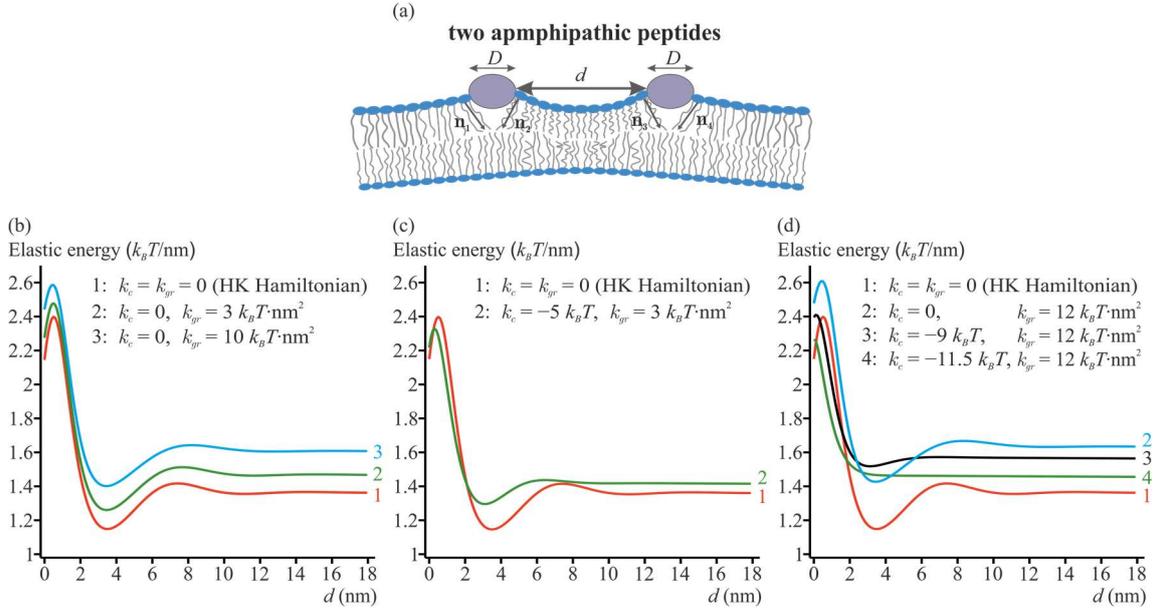

FIG. 2. (Color online) Physical consequences of the new contributions, $k_c \mathbf{T} \cdot (\nabla \tilde{K})$ and $\frac{k_{gr}}{2}(\nabla \tilde{K})^2$, to the elastic energy of lipid membranes, illustrated by an example of a membrane-mediated interaction of two amphipathic peptides located in the same monolayer of a lipid membrane. The figures show the dependence of the elastic energy of the bilayer on the distance $d$ between the peptides. (a) Schematic representation of two amphipathic peptides of diameter $D = 1.3$ nm, inducing a jump in the boundary directors, $n_{2x} - n_{1x} = n_{4x} - n_{3x} = -0.8$. (b) The energy profile obtained within the framework of the HK Hamiltonian (red curve) in comparison with the profiles obtained using Eq. (16) Hamiltonian with different values of $k_{gr}$ and fixed $k_c = 0$. (c) The energy profile obtained within the framework of the HK Hamiltonian (red curve) in comparison with the profile obtained using Eq. (16) Hamiltonian with theoretically estimated values of $k_c = -5\ k_B T$ and $k_{gr} = 3\ k_B T \cdot \text{nm}^2$ (green curve). (d) The energy profile obtained within the framework of the HK

Hamiltonian (red curve) in comparison with the profiles obtained using Eq. (16) Hamiltonian with different values of $k_c$ and fixed $k_{gr}$ = 12 $k_BT·nm^2$.

Thus, we showed that $k_c$ and $k_{gr}$ significantly quantitatively alter the amphipathic peptides interaction profile, while the qualitative features of the profile remain for the wide range of parameters. This indicates that the usage of the new approach is necessary for a quantitative description of the membrane-mediated peptide interaction. For example, the existence of the potential well (global energy minimum) at the distance of 4 nm between two peptides, predicted within the framework of the HK Hamiltonian, might provide an explanation for their cooperation and pore formation in membranes, under the assumption that pores are formed in a highly stressed region between two peptides next to each other [88]. However, as shown in Fig. 2, at certain values of moduli $k_c$ and $k_{gr}$ in corrected functional (16), this energy minimum disappears, and therefore the cooperative assembling of two peptides can be hindered. In fact, there are experiments [89] showing a fivefold difference in leakages, induced by amphipathic peptide GALA, of fluorescent probes from bilayer vesicles formed of POPC (1-palmitoyl-2-oleoyl-sn-glycero-3-phospho-choline) and DOPC (1,2-dioleoyl-sn-glycero-3-phosphocho-line); the difference cannot be explained by the disparity in binding affinities of the peptide to the membranes. POPC and DOPC have approximately the same elastic model parameters: the hydrophobic thickness [90,91], tilt and bending moduli [92], along with close spontaneous curvatures [93]. Thus, peptides' interaction profiles should be similar in both lipids if they are described by HK's Hamiltonian. We hypothesize that the explanation of these experiments lies in the fact that POPC and DOPC have different values of moduli $k_c$ and $k_{gr}$ which in turn lead to the alteration of the GALA peptides pairwise interaction profile. As DOPC has lower leakage than POPC, the potential well in DOPC membranes should be lower than in POPC ones (see Fig. 2). This results in a shorter time spent by peptides being close to each other and, hence, in a reduced rate of the cooperative formation of pores by the peptides.

We point out that, in the case of nonzero tilt-curvature coupling modulus $k_c$ and zero curvature gradient $k_{gr}$ (the TD Hamiltonian), the variational problem leading to the Euler-Lagrange equations is ill-posed. In this case, one of the roots of the characteristic polynomial of Eq. (D2) is purely imaginary, giving rise to oscillating deformation modes, which are energetically unbounded from below and because of which the energy cannot be integrated over an infinite interval. However, it is possible to reformulate the problem, considering the peptides in a box with periodic boundary conditions. Nevertheless, as expected, if $k_c \neq 0$ and $k_{gr} = 0$, the membrane is unstable: the oscillations are energetically unbounded from below. To be specific, we provide a particular example: staying within the one-dimensional case, we consider a box of length 30 nm with periodic boundary conditions and two peptides, the distance between which is equal to 14.2 nm (the exact values of the chosen parameters are not important). The analysis shows that the energy of this configuration per unit length along the axis of the translational symmetry scales as $aA_0^2 + bA_0 + c$, where $A_0$ is the amplitude of the oscillations, given in nanometers, and $a$, $b$, $c$ are known coefficients ($a \approx -140$ $k_BT/\text{nm}^3$, $b \approx 6.7 \times 10^{-3}$ $k_BT/\text{nm}^2$, $c \approx 1.4$ $k_BT/\text{nm}$ at $k_c = -5$ $k_BT$, $k_{gr} = 0$ $k_BT\cdot\text{nm}^2$, and the rest of the membrane elastic parameters being the same as used in Fig. 2). Since $a > 0$, it means that the oscillations are unbounded from below: the energy approaches minus infinity as $A_0$ increases (see Fig. 3). If we now consider the case of two amphipathic peptides in an infinite membrane, separated by the same distance of 14.2 nm as in Fig. 3, the energy nevertheless does not have a lower bound. As has already been said, the arising oscillations cannot be integrated over an infinite interval. However, if we set the amplitude of the oscillations to zero at infinity as a boundary condition, the oscillations in the region between two peptides still exist and the energy scales as $a'A_0^2 + b'A_0 + c'$, where $a' \approx -85$ $k_BT/\text{nm}^3$, $b' \approx 4.3 \times 10^{-3}$ $k_BT/\text{nm}^2$, $c' \approx 1.1$ $k_BT/\text{nm}$, and therefore, since $a' < 0$, when $A_0$ approaches infinity, the energy tends to minus infinity, implying that the membrane is unstable. Thus, since the energy does not have a lower bound, the variational problem is meaningless in the case of $k_c \neq 0$ and $k_{gr} = 0$.

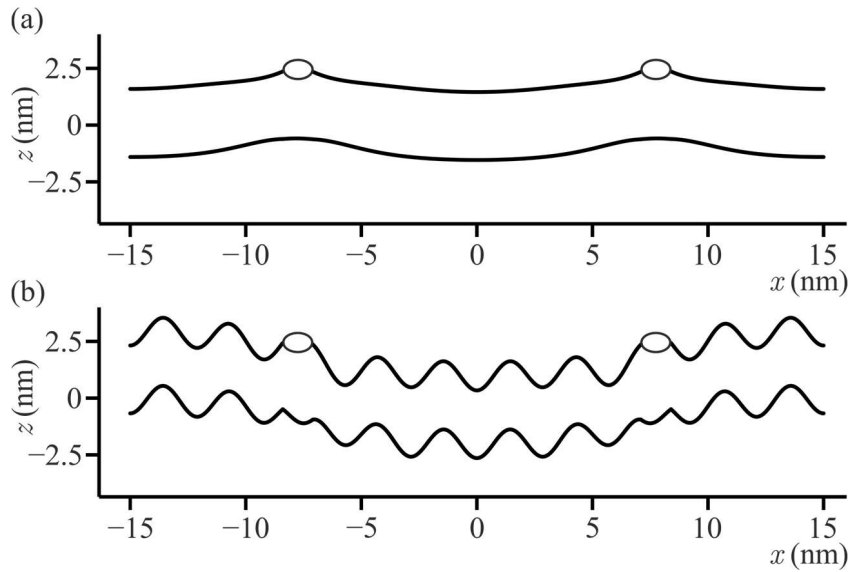

FIG. 3. The lipid membrane is unstable if tilt-curvature coupling modulus $k_c$ is nonzero, while curvature gradient modulus $k_{gr}$ is zero. The instability is caused by membrane oscillations which are energetically unbounded from below. The figure shows two amphipathic peptides (black ellipses) embedded in a box of length 30 nm with periodic boundary conditions. The distance between the peptides is 14.2 nm, and the analytically calculated shapes of the dividing surfaces of the upper and lower monolayers are shown. a) $k_c = -5$ $k_BT$, $k_{gr} = 3$ $k_BT\cdot\text{nm}^2$ (as estimated in Sec. IV.B).

b) $k_c = -5\ k_BT$, $k_{gr} = 0\ k_BT \cdot \text{nm}^2$, amplitude $A_0$ of the oscillations is 0.5 nm. The rest of the membrane elastic parameters are the same as in Fig. 2. In panel (a) the membrane is stable, while in panel (b) the membrane is unstable due to the arising oscillations: energy $w$ of the membrane inside the box scales as $w = aA_0^2 + bA_0 + c$, where $a \approx -140\ k_BT/\text{nm}^3$, $b \approx 6.7 \times 10^{-3}\ k_BT/\text{nm}^2$, $c \approx 1.4\ k_BT/\text{nm}$ (the energy is considered per unit length along the axis of symmetry), implying that when oscillations amplitude $A_0$ increases, the energy of the membrane inside the box approaches minus infinity.

In this paper, we have concentrated only on the membrane-mediated interaction of amphipathic peptides and compared it with previous results. It however does not limit possible implications of new energy terms when applied to other problems, such as the interaction of transmembrane proteins [79–82] or hydrophobic inclusions [83]. In addition, membrane inclusions may induce membrane softening [94–100]. Among the mechanisms of such a phenomenon might be the concentration-curvature coupling [101] or the anisotropy of inclusions [100,102]. Also, the tilt deformation mode might lead to membrane softening [60]. Given that the new energy terms as well as the tilt mode correspond to the transverse shear deformation, they may also modulate this effect.

### E. Fluctuation spectra

A fluctuation analysis is widely used in molecular dynamics simulations for estimations of membrane elastic moduli [14,15,106–112,16,17,47,48,73,103–105]. In this section, we provide theoretical results for the director fluctuations analyzed within the framework of the new elastic energy functional and compare them with those obtained by TD [47] and used in Ref. [48] to fit the fluctuation data.

But before that, we would like to carefully consider the approach which is used to analyze tilt and director fluctuations. Initially, such an analysis was introduced in Ref. [14] for tilt fluctuations and in Ref. [17] for director fluctuations. The approach consists in prescribing Boltzmann probability distribution $P \propto \exp(-F(\mathbf{g})/k_BT)$ to find a system in a state $\mathbf{g}$, where $F$ is the elastic free energy of a bilayer. In Ref. [14], the HK Hamiltonian without the twist term is used for $F$, whereas in Ref. [17] $F$ is equivalent to the HK Hamiltonian in the Monge gauge amended by the twist term. We recall that the HK Hamiltonian as well as Hamiltonian (20) is the equilibrium free energy while director $\mathbf{n}$ and tilt $\mathbf{T}$ are timed-averaged quantities. But in simulation analyses, instantaneous values of $\mathbf{n}$ and $\mathbf{T}$ are measured. However, it is not apparent whether the dynamics of individual lipid molecules is in compliance with the same free energy $F$, which reflects time-averaged equilibrium properties. In other words, it is not obvious whether instantaneous values of various quantities, such as the director or tilt, measured in simulations can replace the mean ones in the energy (20) or HK Hamiltonian to predict the energy of a given state and then the spectrum.

Following [47] and [16], we analyze fluctuations in terms of longitudinal and transverse components of vectors in Fourier space. Using incompressibility condition (A11) up to the first order, we get $H_u - M = h - h^2 \boldsymbol{\nabla} \cdot \mathbf{n}_u / 2 - \alpha_u h$ and $M - H_l = h - h^2 \boldsymbol{\nabla} \cdot \mathbf{n}_l / 2 - \alpha_l h$, where indices $u$ and $l$ correspond to the upper and lower monolayers, respectively, $H$ and $M$ are shapes of the dividing and monolayer interface surfaces, respectively, $\alpha$ is stretching of the dividing surfaces, $h$ is the hydrophobic thickness of undeformed monolayers. We substitute these conditions into (20) and obtain the coupling matrix for the functions { $H_+ = (H_u + H_l)/2$, $H_- = (H_u - H_l - 2h)/2$, $M$, $\mathbf{n}_+ = (\mathbf{n}_u + \mathbf{n}_l)/2$, $\mathbf{n}_- = (\mathbf{n}_u - \mathbf{n}_l)/2$ } in Fourier space, from which we get for longitudinal components of $\mathbf{n}_-$ a rather bulky expression, see Appendix E, Eq. (E1). To compare with the case of no stretching, considered by TD [47], we take the limit $k_A \to \infty$ and get:

$$q^2 \left\langle \left| n_q^{\|} \right|^2 \right\rangle = \frac{k_t k_B T}{2\left[ \left( k_{gr} k_t - k_c^2 \right) q^2 + k_m k_t \right]}, \qquad (30)$$

where $q$ is the wavenumber modulus. Note that $k_{gr} k_t - k_c^2 > 0$ due to stability requirements, and therefore $q^2 \left\langle \left| n_q^{\|} \right|^2 \right\rangle$ is a monotonically decreasing function of $q$. More general expression (E1) is also a monotonically decreasing function of $q$ due to stability requirements (2) (see the proof in Appendix E). In contrast to these predictions, the expression $q^2 \left\langle \left| n_q^{\|} \right|^2 \right\rangle$ increases in simulations at large $q$ [17,48,73,113–115], while at small $q$, it behaves differently depending on algorithms of fluctuation analyses: in Refs. [17,73,113,114] it is constant, in Ref. [115] it decreases and in Ref. [48] it is variable.

Simulations data for $q^2 \left\langle \left| n_q^{\|} \right|^2 \right\rangle$ spectrum were fitted by TD in Ref. [48]. However, this was achieved at the cost of the unstable energy functional. TD's expression for longitudinal fluctuations can be obtained from (30) by setting $k_{gr} = 0$, since there is no term proportional to $\left( \boldsymbol{\nabla} \tilde{K} \right)^2$ in their functional [47]. The same is true for the functional derived in Ref. [48]. Thus, one can see that the spectrum predicted by TD monotonically increases up to the point of divergence $q_0 = \sqrt{k_m k_t} / |k_c|$ and becomes inapplicable at higher $q$. This divergence is the very feature that allows fitting simulation spectra. Thus, so far the longitudinal spectrum of the director has been fitted only with the divergent expression in Ref. [48], but it seems rather artificial since it is achieved by using the unstable energy functional derived by neglecting the second-order term $(\nabla \tilde{K})^2$. In Fig. 4, we demonstrate the comparision between the fluctuation spectrum of $q^2 \left\langle \left| n_q^{\|} \right|^2 \right\rangle$ as predicted by the HK Hamiltonian (Eq. (14) with $k_c = k_{gr} = 0$), TD Hamiltonian (Eq. (14) with $k_c \neq 0$, $k_{gr} = 0$) and the Hamiltonian presented in this work (Eq. (14) with $k_c \neq 0$ and $k_{gr} \neq 0$); for $k_c$ and $k_{gr}$ we use the values estimated in Sec. IV.B ($k_c \approx -5\ k_BT$, $k_{gr} \approx 3\ k_BT \cdot nm^2$, $T = 300$ K), and for $k_m$ and $k_t$ we use the same values as in Sec. IV.D ($k_m = 10\ k_BT$, $k_t = 12\ k_BT/nm^2$, $T = 300$ K).

If $k_{gr} = 0$, then at high $q$-values $\left\langle \left| n_q^{\|} \right|^2 \right\rangle$ becomes negative which implies that the coupling matrix is no longer positive definite, and this causes the instability. The instability can be simply illustrated in the case of a one-dimensional infinite monolayer subjected to periodic boundary conditions with period $L$. In this case, the TD Hamiltonian, per unit length along the axis of the translational symmetry, is $\int dx \left( \frac{1}{2} k_m n'^2 + \frac{1}{2} k_t t^2 + k_c t \cdot n'' \right)$, where $t$ and $n$ are projections of the tilt and director onto the $x$-axis which is parallel to the undeformed monolayer. This expression can be rewritten as $\int dx \left( \frac{1}{2} k_m n'^2 + \frac{1}{2k_t}(k_t t + k_c n'')^2 - \frac{k_c^2}{2k_t} n''^2 \right)$. If we now choose $k_t t + k_c n'' = 0$ and $n = n_0 \sin(qx)$, where $q = 2\pi m/L$, the energy of the membrane over period $L$ becomes $\left( n_0^2 L / 4k_t \right) q^2 \left( k_m k_t - k_c^2 q^2 \right)$. This energy is negative at $q > q_0 = \sqrt{k_m k_t} / |k_c|$ and approaches minus infinity as $q \to +\infty$, implying the instability of the membrane. Because the TD Hamiltonian is unbounded from below, i.e. it is unstable, solving Euler-Lagrange equations for such a Hamiltonian is meaningless. We point out that the application of the lateral tension, $\sigma \cdot H_u'^2 / 2 + \sigma \cdot H_l'^2 / 2$, does not help to recover the stability of the membrane. In fact, Eq. (30) in the case of $k_{gr} = 0$ and nonzero $\sigma$, becomes: $q^2 \left\langle \left| n_q^{\|} \right|^2 \right\rangle = \dfrac{q^2 (k_t + \sigma) k_B T}{2 \left[ -k_c^2 q^4 + \left( k_m k_t + (k_m - 2k_c) \sigma \right) q^2 + \sigma k_t \right]}$, which is nevertheless negative at large $q$: the numerator is always positive, as tilt modulus $k_t$ and tension $\sigma$ are positive, while the denominator is negative at large $q$ because, although $\left(k_m k_t + \left(k_m - 2k_c\right)\sigma\right)q^2 + \sigma k_t$ is always positive ($k_m > 0$ and $k_c < 0$), $q$ to the fourth power has negative coefficient $-k_c^2$.

An alternative explanation of such discrepancies involves a microscopic noise [17,113], which is proposed as the cause of the spectrum increase at large $q$ in simulations. At the same time, at small $q$, the spectra of longitudinal components behave differently depending on algorithms of the analysis of fluctuations, which can be consistent with the predicted monotonic decrease of spectrum (E1).

In order to assess the dependence of function (E1) on $q$, we use the values of moduli, theoretically estimated in Sec. IV.B: $k_c \approx -5\ k_B T$, $k_{gr} \approx 3\ k_B T \cdot \text{nm}^2$, $B \approx -9\ k_B T/\text{nm}$, $C \approx 5\ k_B T \cdot \text{nm}$, and take $k_m = 10\ k_B T$, $k_A = 30\ k_B T/\text{nm}^2$ [85] ($T = 300$ K) and $k_t = 12\ k_B T/\text{nm}^2$ [27]. At these values of the moduli, function (E1) depends only on bending-compression coupling modulus $A$, which lies within the interval $\left(-\sqrt{k_m k_A},\ \sqrt{k_m k_A}\right) \approx \left(-17, 17\right)\ k_B T/\text{nm}$ due to stability requirements. It turns out that the decrease of the function within the range $0 \leq q \leq 1$ is less than 20 % at a rather large range of the values of $A$ (from –9 to 15 $k_B T$/nm). Thus, the decrease of $q^2 \left\langle \left|n_q^\parallel\right|^2 \right\rangle$ is expected to be not large at small values of $q$, when the microscopic noise is not significant. We also point out that

$$q^2 \left\langle \left|n_q^\parallel\right|^2 \right\rangle \xrightarrow[q \to 0]{} \frac{k_B T}{2\left(k_m - A^2/k_A\right)},$$

which coincides with a well-known result of Ref. [17] up to the definition of the bending-stretching coupling modulus $A$.

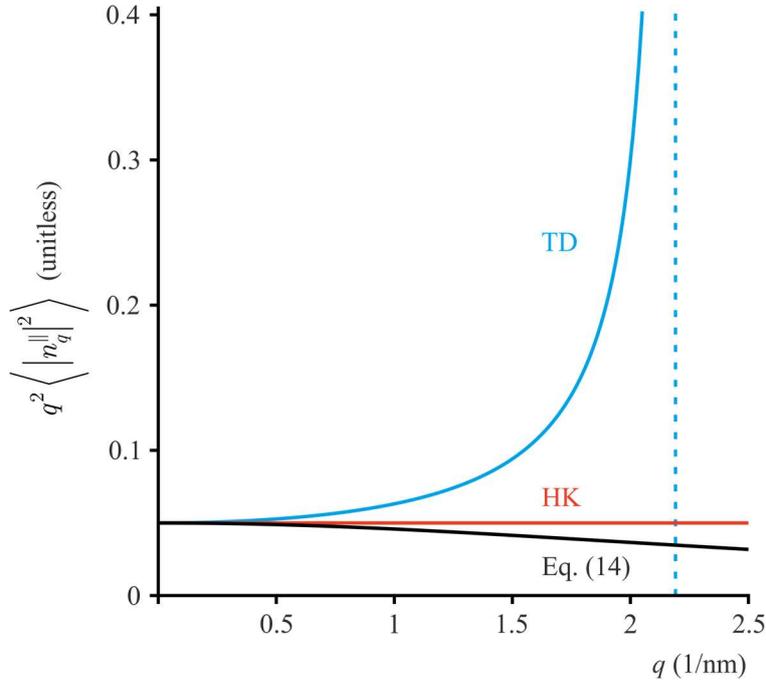

FIG. 4. (Color online) The fluctuation spectrum of $q^2 \langle |n_q^\parallel|^2 \rangle$ as predicted by the HK ($k_c = k_{gr} = 0$, red curve), TD ($k_c \neq 0$, $k_{gr} = 0$, cyan curve) and Eq. (14) ($k_c \neq 0$, $k_{gr} \neq 0$, black curve) Hamiltonians, where $k_c$ and $k_{gr}$ are the tilt-curvature coupling and curvature gradient modulus, respectively. The spectrum is constant in the case of the HK Hamiltonian, monotonically increasing and monotonically decreasing in the case of the TD and Eq. (14) Hamiltonians, respectively. Although the TD Hamiltonian rightly reflects an upward trend observed in simulations at high $q$-values, it is divergent (the corresponding asymptote is depicted as a dashed line). Fixing the divergence leads to the Eq. (14) Hamiltonian, which is monotonically decreasing.

For the transverse component of the director fluctuation, we obtain a constant spectrum $\langle |n_q^\perp|^2 \rangle = \frac{k_B T}{2 k_t}$, whereas in simulations this spectrum is a decreasing function of $q$ [16,17,48,73,113]. To fix this deviation from the theoretical prediction, in Refs. [16,48] the twist contribution $\frac{k_{tw}}{2}(\nabla \times \mathbf{T})^2$, derived by HK in Ref. [27], was added to the energy functionals, which leads to the altered expression: $\langle |n_q^\perp|^2 \rangle = k_B T / 2(k_t + k_{tw} q^2)$. However, in Appendix C, we showed that HK's derivation is incomplete. More importantly, in Sec. III.B we showed that $k_{tw}$ should be equal to zero due to the equivalence of strong and weak lateral fluidity conditions. Therefore, the discrepancy between the theory and simulations remains.

To tackle this issue, we recall the underlying assumptions made to derive energy functional (20). First of all, expression (20) is the equilibrium free energy. It means that $\mathbf{T}, \mathbf{n}, \alpha$ should be considered as time-averaged quantities. Secondly, the corresponding moduli in (20) reflect the energy cost of variations of these time-averaged quantities. Using these two assumptions along with the lateral fluidity of monolayers, we concluded that lateral shear modulus and, therefore, $k_{tw}$ are equal to zero. However, although they are thermodynamically liquid, at small time scales $\sim 0.1$ ns, which appear in fluctuation analyses [17], lateral modes of lipid motion might be hindered. This can impact fluctuations, inducing nontrivial dependence of $\left\langle \left| n_q^\perp \right|^2 \right\rangle$ on $q$. In view of the above, we suggest that the membrane fluctuations, which include the dynamics of individual lipids, cannot be described in detail within the framework of equilibrium free energy functional (20).

### F. Incompressibility assumption in the elastic energy functional

An additional subtle point is worth being considering. The energy functional in (1) is defined on the set of all possible symmetric tensors $\mathbf{U}$, whereas the incompressibility condition restricts $\mathbf{U}$ to the set satisfying the condition $\left| \det(\nabla \mathbf{X}') \right| = 1 \Rightarrow \left| 2\mathbf{U} + \mathbf{1} \right| = 1$. Because of this, the energy also must be defined over this set, but then the Taylor series cannot be written in the form of (1), as the derivatives of $F$ with respect to $u_{ij}$ are not properly defined. This issue can be overcome by extending the energy function to the set of all symmetric tensors [116]. Another approach is to express, for example, $u_{zz}$ through other components of $\mathbf{U}$ from the incompressibility condition, which implies that $u_{zz} = -\left( u_{xx} + u_{yy} \right)$ up to the first order. From the lateral fluidity assumption, it follows that the energy density up to the second order depends only on the combination $\left( u_{xx} + u_{yy} \right)$ rather than on $u_{xx}$ and $u_{yy}$ separately. Therefore, the Taylor series can be written as $F = \sigma_0(z) u_{zz} + \frac{1}{2} E(z) u_{zz}^2 + 4\lambda_5(z) \left( u_{xz}^2 + u_{yz}^2 \right)$. Nevertheless, the final answer for the two-dimensional energy density of the monolayer does not change.

### ACKNOWLEDGMENTS

The work was supported by the Ministry of Science and Higher Education of the Russian Federation and by the grant of the President of the Russian Federation MK-3119.2019.4 . TRG was supported by the Russian Science Foundation grant # 19-74-00152.

# APPENDIX A: KEY RELATIONS

In this section, we provide the derivation of the key formulas, which are used in the main text: for vectors $\mathbf{e}'_i \equiv \nabla_i \mathbf{X}'$, the vector product $\mathbf{e}'_1 \times \mathbf{e}'_2$, $\mathbf{e}'^2$ and the lateral strain $\varepsilon \equiv \|\mathbf{e}'_1 \times \mathbf{e}'_2\| - 1$. We denote by $\nabla_i$ the covariant derivative operator with $i = 1$ or $i = 2$ corresponding to $x$ or $y$, respectively. This operator equals to simple partial derivatives when it acts on scalars and vectors. Besides, the following notations are used: $\mathbf{e}_i = \nabla_i \mathbf{X}$, $\mathbf{e}'_i = \nabla_i \mathbf{X}'$, $g_{ij} = \mathbf{e}_i \cdot \mathbf{e}_j$, $\mathbf{e}^m = g^{mk} \mathbf{e}_k$, $\mathbf{T} = T^k \mathbf{e}_k = T_m \mathbf{e}^m$ with $\mathbf{e}_i$, $\mathbf{e}'_i$ being the surface basis vectors; $g_{ij}$, $g^{mk}$ — the metric tensor and its inverse; $T^k, T_m$ — the components of the tilt vector. In addition, the following equations hold: $\nabla_i \mathbf{N} = K_i^k \mathbf{e}_k$, $\nabla_i \mathbf{e}_k = -K_{ik} \mathbf{N}$, where $K_{ik}$ is the curvature tensor [52]. Following the definitions given by HK and TD, we refer to $\tilde{K}_{ik} \equiv K_{ik} + \nabla_i T_k$ as the effective curvature tensor.

The expression for $\mathbf{e}'_i$ was given in Ref. [47]:

$$\mathbf{e}'_i = [\delta_i^k + \zeta \tilde{K}_i^{\ k} + T^k \nabla_i \zeta] \mathbf{e}_k - [\zeta T^k \tilde{K}_{ik} - \nabla_i \zeta] \mathbf{N} = A_i^{\ k} \mathbf{e}_k - B_i \mathbf{N}, \tag{A1}$$

where we defined $A_i^{\ k}$ and $B_i$ as:

$$\begin{aligned} A_i^{\ k} &= [\delta_i^k + \zeta \tilde{K}_i^{\ k} + T^k \nabla_i \zeta], \\ B_i &= [\zeta T^k \tilde{K}_{ik} - \nabla_i \zeta]. \end{aligned} \tag{A2}$$

In the case of $\sqrt{g} = 1$, the vector product $\mathbf{e}'_1 \times \mathbf{e}'_2$ can be written as:

$$\mathbf{e}'_1 \times \mathbf{e}'_2 = \frac{1}{2} \varepsilon^{ij} \mathbf{e}'_i \times \mathbf{e}'_j, \tag{A3}$$

where $\varepsilon^{ij}$ is the Levi-Civita tensor. Defined as $\varepsilon^{ij} \equiv \dfrac{\epsilon^{ij}}{\sqrt{g}}$, where $\epsilon^{ij}$ is the Levi-Civita symbol, the Levi-Civita tensor obeys the following identities:

$$\mathbf{e}_i \times \mathbf{e}_j = \varepsilon_{ij} \mathbf{N},\ \mathbf{N} \times \mathbf{e}_i = \varepsilon_{ij} \mathbf{e}^j,\ \varepsilon^{ij} \varepsilon_{kj} = \delta_k^i,\ \varepsilon^{ij} \varepsilon_{ij} = 2,\ \varepsilon^{mn} \varepsilon_{ij} = \delta_i^m \delta_j^n - \delta_i^n \delta_j^m. \tag{A4}$$

Using these properties, we can write:

$$\begin{aligned} \mathbf{e}'_i \times \mathbf{e}'_j &= [A_i^{\ l} \mathbf{e}_l - B_i \mathbf{N}] \times [A_j^{\ m} \mathbf{e}_m - B_j \mathbf{N}] = A_i^{\ l} A_j^{\ m} \mathbf{e}_l \times \mathbf{e}_m - B_i A_j^{\ m} \mathbf{N} \times \mathbf{e}_m \\ &\quad - A_i^{\ l} B_j \mathbf{e}_l \times \mathbf{N} = A_i^{\ l} A_j^{\ m} \varepsilon_{lm} \mathbf{N} - B_i A_j^{\ m} \varepsilon_{mp} \mathbf{e}^p + B_j A_i^{\ l} \varepsilon_{lp} \mathbf{e}^p \\ &= A_i^{\ l} A_j^{\ m} \varepsilon_{lm} \mathbf{N} + (B_j A_i^{\ m} - B_i A_j^{\ m}) \varepsilon_{mp} \mathbf{e}^p. \end{aligned} \tag{A5}$$

Hence,

$$
\begin{aligned}
\frac{1}{2}\varepsilon^{ij}\mathbf{e}'_i \times \mathbf{e}'_j &= \frac{1}{2}\varepsilon^{ij} A_i{}^l A_j{}^m \varepsilon_{lm}\mathbf{N} + \frac{1}{2}\varepsilon^{ij}(B_j A_i{}^m - B_i A_j{}^m)\varepsilon_{mp}\mathbf{e}^p = \\
&= \frac{1}{2}\varepsilon^{ij} A_i{}^l A_j{}^m \varepsilon_{lm}\mathbf{N} + \frac{1}{2}(\delta^i_m \delta^j_p - \delta^j_m \delta^i_p)(B_j A_i{}^m - B_i A_j{}^m)\mathbf{e}^p = \\
&= \frac{1}{2}\varepsilon^{ij} A_i{}^l A_j{}^m \varepsilon_{lm}\mathbf{N} + \frac{1}{2}\{(B_p A_m{}^m - B_m A_p{}^m) - (B_m A_p{}^m - B_p A_m{}^m)\}\mathbf{e}^p = \\
&= \frac{1}{2}\varepsilon^{ij} A_i{}^l A_j{}^m \varepsilon_{lm}\mathbf{N} + (B_p A_m{}^m - B_m A_p{}^m)\mathbf{e}^p.
\end{aligned}
\tag{A6}
$$

The terms in front of $\mathbf{N}$ and $\mathbf{e}^p$ in Eq. (A6) can be rewritten, up to the required order, as:

$$\frac{1}{2}\varepsilon^{ij} A_i{}^l A_j{}^m \varepsilon_{lm} = 1 + \zeta\tilde{K} + \zeta^2 \tilde{K}_G + \mathbf{T}\cdot\boldsymbol{\nabla}\zeta, \tag{A7}$$

$$
\begin{aligned}
(B_p A_m{}^m - B_m A_p{}^m) &= ([\zeta T^k \tilde{K}_{pk} - \nabla_p \zeta][2 + \zeta\tilde{K} + T^k \nabla_k \zeta] - \\
&- [\zeta T^k \tilde{K}_{mk} - \nabla_m \zeta][\delta^m_p + \zeta\tilde{K}_p{}^m + T^m \nabla_p \zeta]) = -\nabla_p \zeta,
\end{aligned}
\tag{A8}
$$

where $\boldsymbol{\nabla} \equiv \mathbf{e}^i \nabla_i$ is the surface gradient operator. Therefore,

$$\mathbf{e}'_1 \times \mathbf{e}'_2 \equiv \frac{1}{2}\varepsilon^{ij}\mathbf{e}'_i \times \mathbf{e}'_j = \left(1 + \zeta\tilde{K} + \zeta^2 \tilde{K}_G + \mathbf{T}\cdot\boldsymbol{\nabla}\zeta\right)\mathbf{N} - \nabla_p \zeta \mathbf{e}^p. \tag{A9}$$

By definition, $1 + \varepsilon(\zeta) = \left\|\mathbf{e}'_1 \times \mathbf{e}'_2\right\|$. Therefore, from (A9) we get:

$$\varepsilon(\zeta) = \zeta\tilde{K} + \zeta^2 \tilde{K}_G + \mathbf{T}\cdot\boldsymbol{\nabla}\zeta + \frac{1}{2}(\boldsymbol{\nabla}\zeta)^2. \tag{A10}$$

To obtain the expression for $\zeta(z)$ in the case of nonzero stretching $\alpha$, we note that vector product (A9) should be multiplied by $1 + \alpha$. Substituting this vector product to incompressibility condition $\left|[\mathbf{e}'_1 \times \mathbf{e}'_2]\cdot \mathbf{e}'_3\right| = 1$ and solving for $\zeta(z)$ up to the quadratic order gives the following expression:

$$\zeta(z) = \left(1 - \alpha + \alpha^2 + \frac{1}{2}\mathbf{T}^2\right)z - \frac{1}{2}z^2\tilde{K} + z^2\alpha\tilde{K} + \frac{z^3}{3}\tilde{K}^2 - \frac{z^3}{3}\tilde{K}_G \tag{A11}$$

# APPENDIX B: EXPRESSION FOR A LOCAL TILT T($z$)

Here, we derive the expression for a local tilt $\mathbf{T}(z)$ inside the monolayer in order to provide a geometrical interpretation of Eq. (10) for the transverse shears: $4(u_{xz}^2 + u_{yz}^2) = (\mathbf{T} + \boldsymbol{\nabla}\zeta)^2$. We recall that the tilt vector $\mathbf{T}$ is defined as $\mathbf{T} = \dfrac{\mathbf{n}}{\mathbf{n}\cdot\mathbf{N}} - \mathbf{N}$ with $\mathbf{N}$ being the unit normal vector to the surface $\mathbf{X}'(x,y,0)$ and $\mathbf{n}$ — the director. By definition, $\mathbf{T}(z) = \dfrac{\mathbf{n}}{\mathbf{n}\cdot\mathbf{N}(z)} - \mathbf{N}(z)$, where $\mathbf{N}(z)$ is the normal vector to the surface $\mathbf{X}'(x,y,z)$ with $z$ being fixed. Therefore, two vectors $\mathbf{T}$ and $\mathbf{T}(z)$ coincide, if $z = 0$. Using the definition of the surface normal, we obtain:

$$\mathbf{N}(z) = \frac{1}{\sqrt{g'}}\mathbf{e}_1' \times \mathbf{e}_2' = \frac{1}{\sqrt{g}}\mathbf{e}_1' \times \mathbf{e}_2'\frac{\sqrt{g}}{\sqrt{g'}} = \frac{1}{\sqrt{g}}\mathbf{e}_1' \times \mathbf{e}_2'\frac{1}{1+\varepsilon(\zeta)} = \\ = \frac{1}{2}\varepsilon^{ij}\mathbf{e}_i' \times \mathbf{e}_j'\frac{1}{1+\varepsilon(\zeta)}. \tag{B1}$$

Expressions for $\dfrac{1}{2}\varepsilon^{ij}\mathbf{e}_i' \times \mathbf{e}_j'$ and $\varepsilon(\zeta)$ are given in (A9) and (A10), respectively. Thus, up to the required order:

$$\mathbf{N}(z) = \frac{1}{1+\varepsilon(\zeta)}[(1 + \zeta\tilde{K} + \zeta^2 \tilde{K}_G + \mathbf{T}\cdot\boldsymbol{\nabla}\zeta)\mathbf{N} - \nabla_p\zeta\mathbf{e}^p] = \\ = (1 - \frac{1}{2}(\boldsymbol{\nabla}\zeta)^2)\mathbf{N} - \nabla_p\zeta\mathbf{e}^p. \tag{B2}$$

From the definition of the director we get:

$$\mathbf{n} \approx (1 - \frac{1}{2}\mathbf{T}^2)\mathbf{N} + \mathbf{T} = (1 - \frac{1}{2}\mathbf{T}(z)^2)\mathbf{N}(z) + \mathbf{T}(z) = \\ = (1 - \frac{1}{2}\mathbf{T}(z)^2)\left[(1 - \frac{1}{2}(\boldsymbol{\nabla}\zeta)^2)\mathbf{N} - \nabla_p\zeta\mathbf{e}^p\right] + \mathbf{T}(z), \tag{B3}$$

and hence:

$$\mathbf{T}(z) = \frac{1}{2}\mathbf{N}[(\boldsymbol{\nabla}\zeta)^2 + \mathbf{T}(z)^2 - \mathbf{T}^2] + \mathbf{T} + \nabla_p\zeta\mathbf{e}^p. \tag{B4}$$

From this equation, it follows that up to the second order:

$$\mathbf{T}(z)^2 \approx \mathbf{T}^2 + z^2\mathbf{T}\cdot\boldsymbol{\nabla}\tilde{K} + \frac{1}{4}z^4(\boldsymbol{\nabla}\tilde{K})^2 = (\mathbf{T} + \boldsymbol{\nabla}\zeta)^2. \tag{B5}$$

# APPENDIX C: HK'S FLUIDITY ASSUMPTIONS

In this appendix, we discuss the fluidity assumptions made by HK, showing how the combination $u_{xx} + u_{yy}$ can be expressed via the combination $\tilde{u}^2 \equiv (u_{xx} + u_{yy})^2 - 4(u_{xx}u_{yy} - u_{xy}^2)$ and the lateral strain $\varepsilon \equiv \|\mathbf{e}_1' \times \mathbf{e}_2'\| - 1$.

Using the definition of the strain tensor (3), we can rewrite $4(u_{xx}u_{yy} - u_{xy}^2)$ as:

$$4(u_{xx}u_{yy} - u_{xy}^2) = \mathbf{e}_1'^2 \mathbf{e}_2'^2 - (\mathbf{e}_1' \cdot \mathbf{e}_2')^2 - \mathbf{e}_1'^2 - \mathbf{e}_2'^2 + 1 = \\ = (1+\varepsilon)^2 - \mathbf{e}_1'^2 - \mathbf{e}_2'^2 + 1 = 2\varepsilon + \varepsilon^2 - 2(u_{xx} + u_{yy}). \tag{C1}$$

Hence, if we denote $(u_{xx} + u_{yy})^2 - 4(u_{xz}u_{yy} - u_{xy}^2)$ by $\tilde{u}^2$, we have:

$$\tilde{u}^2 = (u_{xx} + u_{yy} + 1)^2 - (\varepsilon + 1)^2. \tag{C2}$$

From this equation, the combination $u_{xx} + u_{yy}$ can be evaluated as:

$$u_{xx} + u_{yy} \approx \varepsilon + \frac{1}{2}\tilde{u}^2. \tag{C3}$$

From (C3) and energy equation (1), it follows that the modulus corresponding to $\tilde{u}^2$ is $\sigma_l + \frac{\lambda_4}{2}$. Then HK considered separately the conditions of the local fluidity $\sigma_l + \frac{\lambda_4}{2} = 0$ and the global fluidity $\int \left(\sigma_l + \frac{\lambda_4}{2}\right) dz = 0$. However, if $\tilde{u}^2$ is considered as an independent deformation mode, one should require $\sigma_l + \frac{\lambda_4}{2} \geq 0$ for the stability of the functional. Hence, the condition $\int \left(\sigma_l + \frac{\lambda_4}{2}\right) dz = 0$ is equivalent to $\sigma_l + \frac{\lambda_4}{2} = 0$. Furthermore, in HK's equation, $\tilde{u}^2 = z^2 \left[\tilde{K}^2 + (\nabla \times \mathbf{T})^2 - 4\tilde{K}_G\right]$, (where $\nabla \times \mathbf{T} \equiv \varepsilon^{ij} \nabla_i T_j$ with $\varepsilon^{ij}$ being the Levi-Civita tensor) the second-order terms, which generally should be included, are missed. Indeed, we can write $\tilde{u}^2$ as:

$$\begin{aligned}
\tilde{u}^2 &= (u_{xx} + u_{yy})^2 - 4(u_{xx}u_{yy} - u_{xy}^2) = (u_{xx} - u_{yy})^2 + 4u_{xy}^2 = \frac{1}{4}4(u_{xx} - u_{yy})^2 + 4u_{xy}^2 \approx \\
&\approx \frac{1}{4}\left[(g_{11} - g_{22}) + 2z\left(\tilde{K}_{11} - \tilde{K}_{22}\right)\right]^2 + \left[g_{12} + z\left(\tilde{K}_{12} + \tilde{K}_{21}\right)\right]^2 = \\
&= \frac{1}{4}(g_{11} - g_{22})^2 + g_{12}^2 + \\
&+ z\left\{(g_{11} - g_{22})(\tilde{K}_{11} - \tilde{K}_{22}) + 2g_{12}\left(\tilde{K}_{12} + \tilde{K}_{21}\right)\right\} + \\
&+ z^2\left\{\left(\tilde{K}_{11} - \tilde{K}_{22}\right)^2 + \left(\tilde{K}_{12} + \tilde{K}_{21}\right)^2\right\}.
\end{aligned} \tag{C4}$$

From this equation, it follows that the terms multiplied by $z$ to the zeroth power and to the first power also have the second order of smallness.

APPENDIX D: EULER-LAGRANGE EQUATIONS FOR THE ONE-DIMENSIONAL CASE

In this appendix, we derive and solve Euler-Lagrange equations for functional (16) for the one-dimensional case. We consider a bilayer which is translationally symmetric along the $y$-axis. We choose the direction of the $x$-axis along the bilayer plane, and of the $z$-axis — perpendicular to it. Let $H_u = H_u(x)$ and $H_l = H_l(x)$ denote the shapes of the dividing surfaces of the upper and lower monolayers, $M \equiv M(x)$ is the shape of the membrane mid-surface, $n_u \equiv n_u(x)$ and $n_l \equiv n_l(x)$ are projections of the upper and lower director fields ($\mathbf{n}_u$ and $\mathbf{n}_l$) onto the $x$-axis. We now note that $\nabla \cdot \mathbf{n}_u \approx \dfrac{d}{dx} n_u$ and $\nabla \cdot \mathbf{n}_l \approx \dfrac{d}{dx} n_l$. Incompressibility condition (9) up to the first order can be written as $H_u(x) - M(x) = h - \dfrac{1}{2} h^2 \dfrac{d}{dx} n_u$ and $H_l(x) - M(x) = -h + \dfrac{1}{2} h^2 \dfrac{d}{dx} n_l$, where $h$ is the thickness of the hydrophobic part of the monolayer. We also add lateral tension terms to the energy functional: $\sigma \cdot \left( \sqrt{1 + \dfrac{d}{dx} H_u^{\,2}} - 1 \right) \approx \dfrac{1}{2} \sigma \cdot \left( \dfrac{d}{dx} H_u \right)^2$ and $\sigma \cdot \left( \sqrt{1 + \dfrac{d}{dx} H_l^{\,2}} - 1 \right) \approx \dfrac{1}{2} \sigma \cdot \left( \dfrac{d}{dx} H_l \right)^2$ per unit length along the $y$-axis. The tilt fields of the upper and lower monolayers are $T_u \approx N_u - \dfrac{d}{dx} M + \dfrac{1}{2} h^2 \dfrac{d^2}{dx^2} n_u$ and $T_l \approx n_l + \dfrac{d}{dx} M + \dfrac{1}{2} h^2 \dfrac{d^2}{dx^2} n_l$. We write the bilayer energy per unit length along the $y$-axis as the sum of the respective energy of both monolayers and obtain the following Euler-Lagrange equations:

$$\begin{cases} \left( h^2 - l^2 + 2l_c \right) \dfrac{d^2}{dx^2} n_+ + \left( \dfrac{1}{4} h^2 + l_c h^2 + l_{gr} + \dfrac{1}{4} sh^4 \right) \dfrac{d^4}{dx^4} n_+ + n_+ = 0, \\ -2 \dfrac{d}{dx} M + \left( h^2 - l^2 + 2l_c \right) \dfrac{d^2}{dx^2} n_- + \left( -h^2 s - h^2 - 2l_c \right) \dfrac{d^3}{dx^3} M \\ + \left( \dfrac{1}{4} h^4 + l_c h^2 + l_{gr} + \dfrac{1}{4} sh_u^2 \right) \dfrac{d^4}{dx^4} n_- + n_- = 0, \\ \dfrac{d}{dx} n_- + (-2 - 2s) \dfrac{d^2}{dx^2} M + \left( \dfrac{1}{2} h^2 + l_c + \dfrac{1}{2} sh^2 \right) \dfrac{d^3}{dx^3} n_- = 0, \end{cases} \quad (D1)$$

where $n_+ = n_u + n_b$, $n_- = n_u - n_b$, $l = \sqrt{k_m / k_t}$, $l_c = k_c / k_t$, $l_{gr} = k_{gr} / k_t$, $s = \sigma / k_t$. From here, we obtain the following equation for $M$:

$$\frac{d^6}{dx^6}M - \frac{-l^2 s + 2l_c s - l_u^2}{l_c^2 - l_{gr}s - l_{gr}}\frac{d^4}{dx^4}M - \frac{s}{l_c^2 - l_{gr}s - l_{gr}}\frac{d^2}{dx^2}M = 0. \tag{D2}$$

The solution of this equation has the form:

$$M = C_1 e^{d_1 x} + C_2 e^{d_2 x} + C_3 e^{d_3 x} + C_4 e^{d_4 x} + C_5 x + C_6, \tag{D3}$$

the constant parameters $d_i$ ($i$ = 1, ..., 4) of which are too bulky to be presented here. Then, it is straightforward to get solutions for $n_+$ and $n_-$ by substituting (D3) into (D1).

To apply the obtained solutions to the case of two interacting amphipathic peptides, we subdivide the membrane into five regions: the bilayer between the peptides, two monolayers below the peptides, and two bilayers lying between the left and right peptide and minus and plus infinity, respectively. These regions are connected in such a way as to provide the continuity of the director field and dividing surfaces. The obtained solutions can be used for the bilayer regions. The Euler-Lagrange equations for the individual monolayers are:

$$\begin{cases} \dfrac{d}{dx}n_l + (1+s)\dfrac{d^2}{dx^2}M + \left(\dfrac{1}{2}h^2 + l_c + \dfrac{1}{2}sh^2\right)\dfrac{d^3}{dx^3}n_l = 0, \\ \dfrac{d}{dx}M + \left(h^2 - l^2 + 2l_c\right)\dfrac{d^2}{dx^2}n_l + \left(\dfrac{1}{2}h^2 + l_c + \dfrac{1}{2}sh^2\right)\dfrac{d^3}{dx^3}M \\ + \left(\dfrac{1}{4}h^4 + l_c h^2 + l_{gr} + \dfrac{1}{4}sh_u^2\right)\dfrac{d^4}{dx^4}n_l + n_l = 0 \end{cases} \tag{D4}$$

Using linear transformations, we obtain the equation for $M$, which coincides with (D2). Substituting corresponding solution (D3) into (D4), it is straightforward to obtain the expression for $n_l$.

APPENDIX E: Proof that $q^2 \left\langle \left| n_q^\| \right|^2 \right\rangle$ is a monotonically decreasing function of $q$

In this section, we prove that the general expression for $q^2 \left\langle \left| n_q^\| \right|^2 \right\rangle$ is a monotonically decreasing function of $q$, where $q$ is the wavenumber modulus. To accomplish this, we will show that the derivative of this function with respect to $q$ is always negative, which follows from the stability requirements. The expression for $q^2 \left\langle \left| n_q^\| \right|^2 \right\rangle$ is:

$$
\begin{aligned}
q^2 \left\langle \left| n_q^\| \right|^2 \right\rangle &= \\
&= \left[ -\left(B^2 + 2k_c k_t\right) q^2 + k_A k_t \right] k_B T \Big/ \Big[ \left( -\left(2C^2 + 4k_c k_{gr}\right) k_t - 2B^2 k_{gr} + 4BC k_c + 4k_c^3 \right) q^4 \\
&\quad + \left( \left( -4AC + 2k_A k_{gr} - 4k_m k_c \right) k_t + 4ABk_c - 2k_m B^2 - 2k_A k_c^2 \right) q^2 + 2k_t \left( k_m k_A - A^2 \right) \Big],
\end{aligned}
\tag{E1}
$$

Firstly, we rewrite (E1) in the following way:

$$
q^2 \left\langle \left| n_q^\| \right|^2 \right\rangle \equiv f(q) = \frac{N_2 q^2 + N_0}{D_4 q^4 + D_2 q^2 + D_0},
\tag{E2}
$$

where $N_2$, $N_0$, $D_4$, $D_2$, $D_0$ denote the corresponding coefficients in (E1). Then, taking the derivative of (E2) with respect to $q$, we get:

$$
\frac{df(q)}{dq} = \frac{2u^{1/2} \left( -D_4 N_2 u^2 - 2D_4 N_0 u + D_0 N_2 - D_2 N_0 \right)}{\left( D_4 u^2 + D_2 u + D_0 \right)^2},
\tag{E3}
$$

where $u = q^2$. The discriminant of the quadratic polynomial in brackets in the numerator of (E2), after substitution of $N_2$, $N_0$, $D_4$, $D_2$, $D_0$ in terms of moduli, can be rewritten in the following form:

$$
\text{Disc} = -2k_t \left( AB^2 + 2A k_c k_t - B k_a k_c + C k_a k_t \right)^2 \Delta,
\tag{E4}
$$

where $\Delta$ is the determinant of the quadratic form in $\mathbf{T}$, $\boldsymbol{\nabla}\tilde{K}$, $\boldsymbol{\nabla}\alpha$, which corresponds to the transverse shear deformation:

$$
\begin{aligned}
\int \frac{1}{2} \lambda_T(z) [\mathbf{T} - z\boldsymbol{\nabla}\alpha - \frac{1}{2} z^2 \boldsymbol{\nabla}\tilde{K}]^2 \, dz &= \\
&= \frac{1}{2} k_{t,m} \mathbf{T}^2 + k_c \mathbf{T} \cdot (\boldsymbol{\nabla}\tilde{K}) + \frac{k_{gr}}{2} (\boldsymbol{\nabla}\tilde{K})^2 + B\mathbf{T} \cdot \boldsymbol{\nabla}\alpha - k_c (\boldsymbol{\nabla}\alpha)^2 + C\boldsymbol{\nabla}\alpha \cdot (\boldsymbol{\nabla}\tilde{K}).
\end{aligned}
\tag{E5}
$$

As $\lambda_T(z) \geq 0$ due to the stability requirements, this integral is always positive as well as the corresponding quadratic form in $\mathbf{T}$, $\nabla \tilde{K}$ and $\nabla \alpha$ in the second line of (E5) (we do not consider the unphysical case of $\lambda_T(z) \equiv 0$). Therefore, $\Delta$ is positive and discriminant (E4) is negative. At the same time, $-D_4 N_2$ — the coefficient of $u$ to the second power in the numerator of (E3) — equals $\left(B^2 + 2k_c k_t\right)\Delta/4 < 0$, as $-B^2 - 2k_c k_t > 0$ due to it being one of the principal minors of quadratic form (E5). Hence, $\dfrac{df(q)}{dq} < 0$, and, consequently, $f$ is a monotonically decreasing function of $q$.